\documentclass[nofootinbib,superscriptaddress]{revtex4-1}
\usepackage[utf8]{inputenc}
\usepackage{amsmath}
\usepackage{graphicx}
\usepackage{float}
\usepackage[T1]{fontenc}
\usepackage{mathtools}  % loads �amsmath�
\usepackage{amssymb}
\usepackage{color}
\usepackage{cancel}
\usepackage{framed}
\usepackage{comment}
\usepackage{mathtools}
\usepackage{hyperref}
\usepackage{chngcntr}
\usepackage[normalem]{ulem}
\usepackage[flushleft]{threeparttable}
\usepackage{multirow}
\usepackage{listings}
\usepackage[table]{xcolor}
\usepackage{hhline}
\usepackage[usestackEOL]{stackengine}
\edef\tmp{\the\baselineskip}
\setstackgap{L}{\tmp}

\counterwithin{equation}{section}
\newcommand{\nn}{\nonumber}

\begin{document}
% Don't want date printed
% Make title large and bold
\title{\Large\bfseries Testing multiflavored ULDM models with SPARC}

\author{Lauren Street}
\email{streetlg@mail.uc.edu}
\affiliation{Fermi National Accelerator Laboratory, Batavia, IL 60510, USA}
\affiliation{Department of Physics, University of Cincinnati, Cincinnati, OH 45221 USA}
\author{Nickolay Y. Gnedin}
\affiliation{Fermi National Accelerator Laboratory, Batavia, IL 60510, USA}
\affiliation{Kavli Institute for Cosmological Physics, The University of Chicago, Chicago, IL 60637 USA}
\affiliation{Department of Astronomy and Astrophysics, The University of Chicago, Chicago, IL 60637 USA}
\author{L.C.R. Wijewardhana}
\affiliation{Department of Physics, University of Cincinnati, Cincinnati, OH 45221 USA}

\begin{abstract}
We perform maximum likelihood estimates (MLEs) for single and double flavor ultralight dark matter (ULDM) models using the Spitzer Photometry and Accurate Rotation Curves (SPARC) database.  These estimates are compared to MLEs for several commonly used cold dark matter (CDM) models.  By comparing various CDM models we find, in agreement with previous studies, that the Burkert and Einasto models tend to perform better than other commonly used CDM models.  We focus on comparisons between the Einasto and ULDM models and analyze cases for which the ULDM particle masses are: free to vary; and fixed.  For each of these analyses, we perform fits assuming the soliton and halo profiles are: summed together; and matched at a given radius.  When we let the particle masses vary, we find a negligible preference for any particular range of particle masses, within $10^{-25}\,\text{eV}\leq m\leq10^{-19}\,\text{eV}$, when assuming the summed models.  For the matched models, however, we find that almost all galaxies prefer particles masses in the range  $10^{-23}\,\text{eV}\lesssim m\lesssim10^{-20}\,\text{eV}$. For both double flavor models we find that most galaxies prefer approximately equal particle masses.  We find that the summed models give much larger variances with respect to the soliton-halo (SH) relation than the matched models.  When the particle masses are fixed, the matched models give median and mean soliton and halo values that fall within the SH relation bounds, for most masses scanned.  When the particle masses are fixed in the fitting procedure, we find the best fit results for the particle mass $m=10^{-20.5}\,\text{eV}$ (for the single flavor models) and $m_1=10^{-20.5}\,\text{eV}$, $m_2=10^{-20.2}\,\text{eV}$ for the double flavor, matched model.  We discuss how our study will be furthered using a reinforcement learning algorithm.
\end{abstract}

\maketitle

\section{Introduction}
\label{sec:intro}

A persistent problem in physics is the physical nature of dark matter (DM).  A popular candidate is cold dark matter (CDM) which is thought to envelope galaxies far beyond the reaches of baryonic matter.  On galactic scales, the presence of CDM is thought to be the cause of flat rotation curves at large radii.  However, past CDM-only simulations resulted in galactic halo profiles (NFW profiles \cite{1996ApJ...462..563N,1997ApJ...490..493N,2010MNRAS.402...21N}) that tended to be poor fits to the density profiles of low mass and low surface brightness galaxies; a problem which has commonly become known as the ``cusp-core'' problem.  These galaxies tended to have more cored profiles, such as the Burkert \cite{Burkert:1995yz} profile; which are constant near small radii and asymptote to the NFW profile for large radii.  It has been recently suggested that the Burkert profile is a better fit for larger galaxies as well, compared to the NFW profile \cite{10.1093/mnras/stx1384}. Another recent study \cite{2021} has shown that the Einasto profile is a better fit to the galaxies in the Spitzer Photometry and Accurate Rotation Curves (SPARC) catalog \cite{Lelli:2016zqa} than the NFW profile.  This suggests that the CDM only simulations that resulted in NFW profiles do not give an adequate picture of galactic DM halos.  However, it has also been recently shown that simulated halos of CDM over a large range of masses can be fit well by the Einasto profile \cite{Wang:2019ftp}.  In this case, the ``cusp-core'' problem of the NFW profile can be resolved by noting that the cored Einasto profile can also be used to model simulated CDM only halos.

While the cusp-core problem is now generally considered to be resolved, there is another problem with the traditional CDM only halo model, i.e. the NFW model, known as the ``diversity'' problem \cite{Bullock:2017xww}.  This describes the trend for galaxies with similar maximum circular velocities to exhibit a wide range of inner circular velocity profiles; a trend which is poorly modeled by the NFW profile.  However, it has been shown that both modified Newtownian dynamics (MOND) models \cite{1983ApJ...270..365M,1983ApJ...270..371M} as well as self-interacting dark matter (SIDM) models can be well fit to the diverse ranges of inner profiles \cite{Kamada:2016euw,Ren:2018jpt,Kahlhoefer:2019oyt}.  In \cite{Ren:2018jpt}, it was shown that SIDM can be fit well to many of the galaxies in the SPARC catalog while also reproducing the concentration mass relation (CMR) \cite{Dutton:2014xda}, the abundance matching relation (AMR)\cite{2013,2012}, the baryonic Tully-Fisher relation (BTFR) \cite{2015}, stellar synthesis models \cite{2014}, and the gravitational radial acceleration relation (RAR) \cite{2016}.  It has also been recently suggested that hadronically interacting DM (HIDM) models tend to fit the SPARC catalog galaxies better than traditional SIDM models  \cite{2021}.

It is natural to think that baryonic matter and DM affect each other throughout the evolution of galaxies.  In fact, recent hydrodynamical simulations suggest that the behavior and shape of the cores of galaxies depend on the ratio of stellar and DM masses ($M_*/M_\text{halo}$) \cite{Oh_2011,10.1093/mnras/stt1891,10.1093/mnras/stu729,10.1093/mnras/stv2101,10.1093/mnras/stv1699,10.1093/mnras/stw3101}.  If this is the case, DM only simulations cannot properly describe the cores of galaxies.  A recent study of simulated CDM halos with baryonic and stellar feedback mechanisms \cite{Chan:2015tna} has shown that, in fact, CDM with baryonic effects results in halos without the ``small-scale'' problems of CDM only halos. A phenomenological density profile that takes into account $M_*/M_\text{halo}$, dubbed ``DC14'' \cite{10.1093/mnras/stt1891,10.1093/mnras/stu729}, has recently been shown to be a much better fit to galactic data than the NFW profile \cite{10.1093/mnras/stw3101}.  It has also been shown that the DC14, as well as other cored profiles, generally give better fits to the galaxies in the SPARC catalog than the NFW profile \cite{Li:2020iib}.

Another popular candidate for DM is the QCD axion which was originally theorized to potentially solve the strong CP problem \cite{Peccei:1977hh,Weinberg:1977ma,Wilczek:1977pj}.  Similar types of particles, termed axion-like-particles (ALPs), arise in string compactification and clockwork theories \cite{Arvanitaki:2009fg,Kaplan:2015fuy}, usually such that many different ALPs are theorized to be in existence.  Modeling QCD axions and ALPs as DM has gained an increase in interest, partly due to the failure to discover weakly interacting massive particles in various searches.  On galactic and cosmological scales, QCD axions and ALPs act similarly to CDM.  ALPs, which are bosonic, can naturally form gravitationally bound structures, commonly called solitons, on astrophysical scales, a subject which has been studied in great detail in the recent past \cite{Kaup:1968zz,Ruffini:1969qy,BREIT1984329,Colpi:1986ye,Seidel:1990jh,Friedberg:1986tq,Seidel:1991zh,LEE1992251,Chavanis:2011zi,Chavanis:2011zm,doi:10.1142/S0217732316500905,Eby:2016cnq,Eby:2017azn,VISINELLI201864,Levkov:2018kau,Eby:2018dat,Braaten:2019knj,Kouvaris:2019nzd,sym12010025,Eby:2019ntd,Eggemeier:2019jsu,Kirkpatrick:2020fwd,Eby:2020ply,Kouvaris:2021phj}.  A subset of ALPs, termed ultra-light DM (ULDM), with masses of $m \sim 10^{-22} \, \text{eV}$ have Compton wavelengths on the order of galactic cores \cite{PhysRevD.28.1243,PhysRevLett.64.1084,Sin:1992bg,Hu:2000ke,Hui:2016ltb,Lee:2017qve,Schive:2014dra,Schive:2014hza,Schwabe:2016rze,Veltmaat:2016rxo,Mocz:2017wlg}.  Because of this, these types of particles were theorized to make up the cores of galaxies, in an attempt to solve the small-scale problems, at the time, of CDM \cite{PhysRevD.40.2524,PhysRevD.50.3650,PhysRevD.50.3655,PhysRevD.62.104012,Schunck:1999zu,Amaro-Seoane:2010pks,PhysRevD.81.044031,Weinberg:2013aya,Marsh:2015wka}.  However, as noted previously, the small-scale problems of CDM tend to disappear when taking into account the relationship of baryonic matter and CDM in galaxies.

While ULDM with masses $m \sim 10^{-22} \, \text{eV}$ can potentially form structures on the order of galactic cores, single flavor models that consider these masses have increasingly become constrained \cite{Irsic:2017yje,Leong:2018opi,Bar:2018acw,Bar:2019bqz,Safarzadeh:2019sre,Schutz:2020jox,Benito:2020avv,Broadhurst:2018fei,2021ApJ...913...25C,Bar:2021kti}.  There have been numerous, recent analyses constraining the ULDM mass from galactic data.  Simulations of collapsing ULDM halos suggest a relationship between the mass of the soliton that forms in a galaxy, and the properties of its host halo, termed the soliton-halo (SH) relation.  In \cite{Bar:2018acw}, it was shown that the SH relation implies that the maximum circular velocity of a soliton should be of the same order as the maximum circular velocity of its host halo.  From this implication, the authors showed that low-surface brightness galaxies in the SPARC catalog disfavor the soliton-host halo relation for ULDM masses of $m \sim 10^{-21} - 10^{-22} \, \text{eV}$.  The authors of \cite{Bar:2019bqz} extended this analysis by including external gravitational potentials in order to understand the effect of baryons on the formation and structure of the soliton and host halo.  The results of \cite{Bar:2018acw} were further confirmed with the analysis of \cite{Bar:2019bqz} and another study \cite{Bar:2021kti} that we will describe in more detail later.

Other previous analyses have shown that ULDM masses $m \gtrsim \, 10^{-22}\,\text{eV}$ tend to give good fits for particular ultra faint dwarf (UFD) satellites \cite{10.1093/mnras/stw1256} while masses $m \lesssim 10^{-22} \, \text{eV}$ tend to give good fits for particular dwarf spheriodals (dSphs) \cite{Marsh:2015wka,10.1093/mnras/stx1941,Schive:2014dra}.  However, the masses that can potentially fit UFDs give poor fits for dSPhs, while the masses that can potentially fit the dSPhs result in UFD masses that are too large.  From the upper constraint on the UFD galaxy masses, a lower constraint of $m > 10^{-21} \, \text{eV}$ can be placed on the ULDM mass \cite{Safarzadeh:2019sre}.  In \cite{2021ApJ...913...25C}, a model independent analysis of the SPARC galaxies was done in which both lower and upper constraints could be placed on the ULDM mass.  It was shown that the most constraining galaxy excluded the mass range of $m = (0.14 - 3.11) \times 10^{-22} \, \text{eV}$.  Finally, the analysis of \cite{Bar:2018acw} was extended by doing a systematic scan over possible ULDM masses \cite{Bar:2021kti}.  In this analysis, a conservative constraint was put on the ULDM mass from SPARC catalog galaxies by finding the maximum possible mass of the soliton.  It was shown that structures composed of ULDM masses in the range $10^{-24} \, \text{eV} \lesssim m \lesssim 10^{-20} \, \text{eV}$ that satisfied the soliton-host halo relation were in tension with the rotation curve data of the SPARC galaxies. This analysis was model-independent in the sense that the authors made no assumptions about the nature of the host halo.  Rather, they only used galaxies that had circular velocities which were overshot by the soliton circular velocity in order to place constraints.  The authors also did a model dependent statisical fit to derive similar constraints using a log-likelihood ratio.  For this fit, they assumed the host halo of the soliton could be described by either the NFW or Burkert profile and added the halo to the soliton in two ways.  The first was by simply adding the two together, assuming both the soliton and halo profiles contribute to the galactic DM density, and the second was by matching the inner soliton profile to the outer halo profile at some transition radius.  In both cases, the statistical analysis further confirmed the model independent constraint for ULDM masses in the range $10^{-24} \, \text{eV} \lesssim m \lesssim 10^{-20} \, \text{eV}$.

It is clear, then, that ULDM masses of $m \lesssim 10^{-20}\,\text{eV}$ are well constrained from galactic data.  However, if one models ULDM as being composed of multiple species, as is natural in string and clockwork theories, these constraints can potentially be evaded or decreased.  It is interesting, then, to consider the types of structures that are formed from multiple species on galactic scales.  There have already been several studies to this effect \cite{Kan:2017uhj,Broadhurst:2018fei,Eby:2020eas,Guo:2020tla}, and one can directly compare the resulting models to galactic data in order to test the validity of galactic DM as multiple species of ULDM.  From a recent analysis \cite{Broadhurst:2018fei}, two main species of ULDM (with masses of $m_1\sim 10^{-22} \, \text{eV}$ and $m_2 \sim 10^{-20}\, \text{eV}$) have been inferred from dSphs and UFDs, respectively.  It was also shown that the inner profile of the Milky Way (MW) can be fit to a combination of solitons each composed of a different mass, with one of the solitons making up the DM components of the MW nuclear star cluster.  The possibility of a third species was also suggested from the analysis of the 47 Tuc globular cluster.  This study assumed that each DM galactic structure (except for the MW) was composed of only one species of ULDM, while allowing different galaxies to be composed of different species of ULDM.  The MW, however, was modeled as being composed of ULDM structures formed from two different species.

Besides multiple flavors, ULDM galatic structures can also be composed of multiple energy eigenstates of a single flavor of ALP \cite{Matos:2007zza,Bernal:2009zy,UrenaLopez:2010ur,Lin:2018whl,Guzman:2019gqc,Street:2021qxl}.  In fact, it has been suggested that these multi-state systems perform better than single-state systems when compared to data \cite{Sin:1992bg,Guzman:2006yc,PhysRevD.68.023511,PhysRevD.53.2236,Matos:2007zza,Bernal:2009zy,UrenaLopez:2010ur}.  We acknowledge that models of ULDM galactic halos composed of multiple energy eigenstates can also potentially fit galactic data, and we leave an analysis of these structures for future work.

It is evident, then, that there are multiple theories of galactic DM, including: CDM modeled with cored profiles, SIDM, and HIDM, each of which can potentially either describe many different types of galaxies or galactic empirical relations, or both.  In this paper, we do not dispute the predictive power of any of these theories.  Rather, we discuss theories of ULDM as galactic DM, since they are interesting alternatives to CDM, SIDM, HIDM, and MOND.  ULDM galactic structures are theorized to have specific signatures and can potentially be searched for using pulsar timing arrays, with multiple flavored models of ULDM having more specific signatures than single flavored models of ULDM \cite{Khmelnitsky:2013lxt,DeMartino:2017qsa,PhysRevD.98.102002,Aoki:2016mtn}.  

We compare ULDM models for both single and multiple flavored cases to commonly used CDM models of galactic DM halos using galactic rotation curves from the SPARC catalog  \footnote{We focus on ULDM models without self-interactions between the particles.  For a common class of ULDM self-interaction potentials, it was shown that self-interactions could be neglected for all galactic halos in the SPARC catalog that were analyzed in \cite{Bar:2018acw}.  We leave an analysis concerning different ULDM self-interaction potentials for future work.}. We find the maximum likelihood parameters by minimization of the chi-square statistic and compare models to each other using the Bayesian information criterion (BIC) statistic, which penalizes models with more parameters.  We also check which models, if any, are in tension with many empirically derived relations, including the CMR, the AMR, the BTFR, stellar synthesis models, and the gravitational RAR.  We also check the SH relation for the ULDM models analyzed.

In Sec. \ref{sec:models}, we describe the galactic DM halo models that will be tested.  In Sec. \ref{sec:fits}, we describe the fitting procedure and we compare our analysis to previous studies in Sec. \ref{sec:discussion}.  We discuss our results and possible future implementations in Sec. \ref{sec:results} and conclude in Sec. \ref{sec:conclusion}. All results obtained in this paper can be found in the publicly available code \cite{Street_DM_halo_models}.  Throughout the text, we use the notation $m_{22} = 10^{-22} \, \text{eV}$.

\section{Dark matter halo models}
\label{sec:models}
Here, we discuss the galactic DM halo models that will be tested against galactic data.  We focus on four different models for CDM halos and the goodness of fit of each model is compared to each other as well as the ULDM models to be discussed later.

\subsection{CDM}
\label{sec:models_CDM}
The halo model that resulted from CDM only simulations is the NFW model \cite{1996ApJ...462..563N,1997ApJ...490..493N,2010MNRAS.402...21N} which has a density profile given by,
\begin{align}\label{eq:rho_NFW}
\rho_\text{NFW}(r) = \frac{\rho_s}{\left(r/r_s\right)\left[1+\left(r/r_s\right)\right]^2},
\end{align}
where $\rho_s$ and $r_s$ are some density and radius scale factors, respectively.  A more phenomenologically motivated model is the Burkert model \cite{Burkert:1995yz}, which has a density profile given by,
\begin{align}\label{eq:rho_Burk}
\rho_\text{B}(r) = \frac{\rho_s}{\left[1+\left(r/r_s\right)\right]\left[1+\left(r/r_s\right)^2\right]}.
\end{align}
Another phenomenologically motivated model is the Einasto model \cite{1965TrAlm...5...87E,10.1111/j.1365-2966.2004.07586.x,Wang:2019ftp} with a density profile given by,
\begin{align}\label{eq:rho_Ein}
\rho_\text{E}(r) = \rho_s \exp \left\{-\frac{2}{\alpha}\left[\left(\frac{r}{r_s}\right)^\alpha - 1\right]\right\}.
\end{align}
where $\alpha$ is taken to be a free parameter.  Finally, a model that takes into account the ratio of stellar to DM mass $\left(M_*/M_\text{halo}\right)$ is the DC14 model \cite{10.1093/mnras/stt1891,10.1093/mnras/stu729} which has a density profile given by,
\begin{align}\label{eq:rho_DC14}
\rho_{\text{DC}14}(r) = \frac{\rho_s}{\left(r/r_s\right)^\gamma\left[1+\left(r/r_s\right)^\alpha\right]^{\left(\beta - \gamma\right)/\alpha}},
\end{align}
where
\begin{align}\label{eq:DC14_params}
\alpha &= 2.94 - \log_{10}\left[\left(10^{X+2.33}\right)^{-1.08}+\left(10^{X+2.33}\right)^{2.29}\right],
\nn \\
\beta &= 4.23 + 1.34 X + 0.26 X^2,
\nn \\
\gamma &= -0.06 - \log_{10}\left[\left(10^{X+2.56}\right)^{-0.68}+\left(10^{X+2.56}\right)\right],
\nn \\
X &= \log_{10}\left(\frac{M_*}{M_\text{halo}}\right).
\end{align}
This above equation is only valid in the range $-4.1 < X < -1.3$.  We constrain one of the fit parameters in order to ensure that $X < -1.3$.

For all CDM halos, we define the concentration, $c_\text{200}$, and virial velocity, $V_\text{200}$, as,
\begin{align}
c_\text{200}\equiv \frac{R_\text{200}}{r_s},
\qquad
V_\text{200} \equiv \sqrt{\frac{M_\text{200}}{M_P^2 R_\text{200}}},
\end{align}
where $M_P \approx 1.22 \times 10^{19} \, \text{GeV}$ is the Planck mass, $R_\text{200}$ is the radius at which the average density is equal to $200$ times the critical density, and $M_{200}$ is mass contained within $R_{200}$ and is commonly called the virial mass.   For each of the density profiles,
\begin{align}
\rho_s &= \frac{M_\text{200}}{4 \pi r_s^3 \left[\ln\left(1+c_\text{200}\right)-\frac{c_\text{200}}{1+c_\text{200}}\right]},
\nonumber \\
M_{200} &= \sqrt{\frac{3}{2 \pi \rho_c}}\frac{M_P^3 V_{200}^3}{20} \qquad r_s = \sqrt{\frac{3}{2 \pi \rho_c}}\frac{M_P V_{200}}{20 c_{200}},
\end{align}
where $\rho_c$ is the critical density of the universe.  As in \cite{10.1093/mnras/stw3101}, we take the Hubble constant to be $H_0 = 73 \, \text{km}/\text{s}/\text{Mpc}$.  For all profiles, we also take the total stellar mass to be,
\begin{align}
M_* \approx \left(\Upsilon_d + \Upsilon_b\right) L,
\end{align}
where $\Upsilon_d$ and $\Upsilon_b$ are the stellar mass-to-light ratios of the disk and bulge, respectively, and $L$ is the total luminosity.

\subsection{ULDM}
\label{sec:models_ALPs}
Now, we discuss the ULDM models that will be fit to data and compared to each other as well as the CDM models.  ULDM structures that resulted in simulations consisted of an inner ULDM soliton core and outer ULDM halo which could be approximated by the NFW profile \cite{Schive:2014dra,Broadhurst:2018fei}.  Both the soliton and halo are composed of the same species of ULDM, while the soliton is in the form of a Bose-Einstein condensate, and the halo is in the form of virialized ULDM.  We focus on two cases:  ULDM composed of a single species; and ULDM composed of two species.  For each of these, we take two possible models of the total galactic DM density profile:  a sum of the soliton and halo density profiles; and the soliton density profile matched to the halo density profile at a particular radius.  The second case is more physical, as this is the behavior that is expected from ULDM simulations \cite{Schive:2014dra,Broadhurst:2018fei,Bar:2018acw,Bar:2021kti}.  However, the first case may be a valid description if one assumes that the ULDM species only make up some portion of the total DM energy density \cite{Bar:2021kti}.  For each analysis, we take the ULDM halo to follow the Einasto profile given by Eq. (\ref{eq:rho_Ein}).  Both the Einasto and Burkert profiles give overall better fits than the DC14 and NFW profiles, for the galaxies analyzed.  We choose to use the Einasto profile instead of the Burkert profile due to its ability to fit simulated halos of CDM  \cite{Wang:2019ftp}.  Finally, for each of these cases, we take the ULDM mass to be: a free parameter in the fitting procedure; and fixed by scanning over particular values.

We analyze the profile of solitons composed of a single species and multiple species of ULDM.  Solitons composed of a single species have the density profile as given in \cite{Schive:2014dra,Broadhurst:2018fei,Bar:2018acw,Bar:2021kti}
\begin{align}\label{eq:dens_psi_s}
\rho_\text{sol}(r) \approx \frac{\rho_c}{\left(1+0.091\left(r/r_c\right)^2\right)^8},
\end{align} 
where
\begin{align}
\rho_c \approx 7 \times 10^9 \left(\frac{M_\text{sol}}{10^{9} M_\odot}\right)^4 \left(\frac{m}{m_{22}}\right)^6 \frac{M_\odot}{\text{kpc}^3},
\end{align}
and
\begin{align}
r_c \approx 0.228 \left(\frac{M_\text{sol}}{10^9 M_\odot}\right)^{-1} \left(\frac{m}{m_{22}}\right)^{-2} \, \text{kpc}, 
\end{align}
with $m$ the mass of the ULDM, $M_{\text{sol}}$ the total mass of the soliton, and $m_{22} = 10^{-22} \, \text{eV}$.  Solitons composed of multiple species of ULDM have density profiles that can be approximated as a sum of the density profiles of single species structures \cite{Broadhurst:2018fei}.  In this analysis, we focus on two flavor models which result in a density profile given by,
\begin{align}\label{eq:dens_psi_m}
\rho_\text{sol}(r) \approx \sum_{i=1}^{2}\rho_{\text{sol},i}(r),
\end{align}
where each of the $\rho_{\text{sol},i}$ is given by Eq. (\ref{eq:dens_psi_s}) and each species can have a different mass $m_i$.  

Eq. (\ref{eq:dens_psi_s}) is valid up to $r \sim 3\, r_c$ \cite{Schive:2014dra}.  For the case in which we take the soliton profile to be matched to the halo profile, we take the transition radius to be $r_t = 3 \, r_c$.  For the single flavored model, the total galactic DM density profile is given by,
\begin{align}\label{eq:rhoDM_matched}
\rho^\text{Matched}_\text{Single}(r) =
\begin{cases}
\rho_\text{sol}(r) & \text{if} \qquad r \leq 3 \, r_c
\\
\rho_\text{halo}(r) & \text{if} \qquad r \geq 3 \, r_c,
\end{cases}
\end{align}
where $\rho_\text{sol}$ is given by Eq. (\ref{eq:dens_psi_s}) and we take $\rho_\text{halo}(r)$ to be given by the Einasto profile (Eq. (\ref{eq:rho_Ein})).  For the double flavored model, the total galactic DM density profile is given by,
\begin{align}\label{eq:rhoDM_matched_2}
\rho^\text{Matched}_\text{Double}(r) =
\begin{cases}
\rho_\text{sol,1}(r)+\rho_\text{sol,2}(r) & \text{if} \qquad r \leq 3 \, r_{c,1}
\\
\rho_\text{halo,1}(r) + \rho_\text{sol,2}(r) & \text{if} \qquad 3 \, r_{c,1} \leq r \leq 3 \, r_{c,2}
\\
\rho_\text{halo,1}(r)+\rho_\text{halo,2}(r) & \text{if} \qquad r \geq 3 \, r_{c,2}.
\end{cases}
\end{align}
We note that there have been no simulations analyzing the collapse of ULDM halos composed of multiple species.  In this case, it is not clear whether the double flavor ULDM structures can be modeled as two solitons each matched to a halo.  However, we choose to extrapolate the results of the single flavor ULDM simulations to double flavor models.  The results of this study can then be compared to any future simulation analyses for double flavor models.  The relation 
\begin{align}\label{eq:matched_relation}
\rho_\text{sol,i}(3 \, r_{c,i}) = \rho_\text{halo,i}(3 \, r_{c,i}),
\end{align}
allows one free parameter to be fixed for the single flavor model and two for the double flavor model.  For the single flavor model, we choose the Einasto halo profile variable $\alpha$ to be fixed, while for the double flavor model we choose each of the Einasto halo profile variables $\alpha_1$ and $\alpha_2$ to be fixed. We choose to solve for $\alpha$ from Eq. (\ref{eq:matched_relation}) since it can be solved analytically and we can use the same halo profile variables when assuming just CDM models (i.e. $c_{200}$ and $V_{200}$).

For the case in which the soliton and halo profiles are summed together, we take the soliton profile to be valid for all $r$, as for $r > 3 \, r_c$, the eigth power in the denominator causes the density profile to fall off rapidly.  For the single flavored model, the total galactic DM density profile is given by,
\begin{align}\label{eq:rhoDM_summed}
\rho^\text{Summed}_\text{Single}(r) = \rho_\text{sol}(r) + \rho_\text{halo}(r).
\end{align}
For the double flavored model, the total galactic DM density profile is given by,
\begin{align}\label{eq:rhoDM_summed_2}
\rho^\text{Summed}_\text{Double}(r) = \sum_{i=1}^2 \left[\rho_\text{sol,i}(r) + \rho_\text{halo,i}(r)\right].
\end{align}

It has been shown from simulations of collapsing halos consisting of a single species of ULDM that the core mass and halo mass follow a scaling relation given by \cite{Schive:2014hza,Bar:2018acw},
\begin{align}\label{eq:core_halo_relation}
M_\text{sol} \approx 1.4 \times 10^9 \left(\frac{m}{m_{22}}\right)^{-1} \left(\frac{M_\text{halo}}{10^{12} M_\odot}\right)^{1/3} \, M_\odot,
\end{align}
for halo masses, $M_\text{halo}$, greater than some minimal mass given by  \cite{Bar:2018acw},
\begin{align}
M_{\text{halo,min}} \sim 5.2 \times 10^7 \left(\frac{m}{m_{22}}\right)^{-3/2} \, M_\odot.
\end{align}
We do not impose this relation, rather we check that the constraint is satisfied after performing fits.  As discussed above, it is unclear whether ULDM halos composed of multiple species will collapse to have the same structure as those composed of single flavor models.  In this case, the SH relation may not even hold for double flavor models.

\section{Rotation curves}
\label{sec:fits}
The SPARC catalog \cite{Lelli:2016zqa} gives, at a given radius from the center of a galaxy, the total observed rotation velocity $V_\text{obs}$, the gas contribution to the rotation velocity $V_\text{gas}$, and the disk and bulge contributions to the rotation velocity, $V_\text{disk}$ and $V_\text{bulge}$ assuming a stellar mass-to-light ratio $\Upsilon_* = 1 M_\odot/L_\odot$.  The contribution of baryonic matter to the total rotation velocity can then be defined as,
\begin{align}
V_\text{bar}(r) \equiv \sqrt{\left|V_\text{gas}(r)\right| V_\text{gas}(r) + \tilde{\Upsilon}_d \left|V_\text{disk}(r)\right| V_\text{disk}(r) + \tilde{\Upsilon}_b\left|V_\text{bulge}(r)\right|V_\text{bulge}(r)},
\end{align}
where $\Upsilon_d = \tilde{\Upsilon}_d \Upsilon_*$ and $\Upsilon_b =  \tilde{\Upsilon}_b \Upsilon_*$ are the stellar mass-to-light ratios of the disk and bulge.  In \cite{Lelli:2016zqa}, the effect of choosing different values for $\tilde{\Upsilon}$ are explored, where the chosen values for the disk and bulge components are $\Upsilon_b = 1.4 \Upsilon_d$.  In our fitting procedure, we take both $\tilde{\Upsilon}_d$ and $\tilde{\Upsilon}_b$ as free parameters.  From certain stellar synthesis models \cite{2014}, the distribution of mass-to-light ratios of the disk and bulge are expected to peak at $\tilde{\Upsilon}_d = 0.5$ and $\tilde{\Upsilon}_b = 0.7$ for surface photometry at $3.6 \, \mu \text{m}$.  In our fits, we take both mass-to-light ratios as free parameters, and check that the resulting distributions peak around the same values.

Assuming spherical symmetry of the DM halo, the DM contribution to the galactic rotation velocity at some distance $r$ from the center of the galaxy is defined as,
\begin{align}\label{eq:VDM}
V_\text{DM}(r) \equiv \sqrt{\frac{4 \pi \int_0^r dr' \, \left(r'\right)^2 \rho_\text{DM}(r')}{M_P^2 \, r}}.
\end{align}
The total observed rotation velocity at a given radius can then be defined as,
\begin{align}\label{eq:Vobs}
V_\text{obs}(r) \equiv \sqrt{V_\text{DM}(r)^2 + V_\text{bar}(r)^2}.
\end{align}

We use LMFIT:  Non-Linear Least-Square Minimization and Curve-Fitting for Python \cite{newville_matthew_2014_11813} to find the maximum likelihood estimation (MLE) by minimization of the chi-square function and Uncertainties: a Python package for calculations with uncertainties \cite{uncertainties} to handle the error calculations.  The chi-square function minimized is given by,
\begin{align}\label{eq:chi_square}
\chi^2 = \sum_{i=1}^{N} \left(\frac{V_\text{model}\left(r_i, \overline{p}\right) - V_i}{\sigma_i}\right)^2.
\end{align}
Here, $V_i$ is the measured total circular velocity and $\sigma_i$ is the error in the measured total circular velocity at the radius $r_i$, while $V_\text{model}\left(r_i, \overline{p}\right)$ is the modeled circular velocity for the parameter set $\overline{p}$. We test the significance of each model compared to each other model using the BIC statistic given by \cite{10.2307/2958889,Liddle:2007fy,Li:2020iib},
\begin{align}
\text{BIC} \equiv -2 \ln(\mathcal{L}_\text{MLE}) + k \ln N,
\end{align}
where $k$ is the number of parameters in the model and $N$ is the number of data points. We take $\mathcal{L}_\text{MLE} \sim \exp(-\chi^2_\text{MLE}/2)$ where $\chi^2_\text{MLE}$ is given by Eq. (\ref{eq:chi_square}) with $\overline{p}_\text{MLE}$ the parameter set which gives the MLE. We find the difference in the BIC statistic between models, $\Delta \text{BIC} = \text{BIC}_{\text{model}_1} - \text{BIC}_{\text{model}_2}$, and use Jeffreys' scale \cite{Jeffreys} to test significance, where $2 < \left|\Delta \text{BIC}\right| \leq 6$ denotes mild evidence, $6 < \left|\Delta \text{BIC}\right| \leq 10$ denotes strong evidence, and $10 < \left|\Delta \text{BIC}\right|$ shows decisive evidence for model 1 (negative $\Delta \text{BIC}$ values) or model 2 (positive $\Delta \text{BIC}$ values).  For $\left|\Delta \text{BIC}\right| \leq 2$, neither model is preferred.  

We choose to compare the BIC statistic as well as the reduced chi-square statistic given by,
\begin{align}\label{eq:red_chi_square}
\chi^2_\nu=\chi^2/(N-k)
\end{align}
where $\chi^2$ is given by Eq. (\ref{eq:chi_square}).  The reduced chi-square tends to treat models similarly when the total number of data points is large compared to the number of model parameters.  However, the BIC statistic tends to be more conservative due to its stricter penalization of models with more parameters.  Therefore, by comparing both the reduced chi-square and BIC for each model analyzed, we can potentially infer if the ULDM models with more parameters are significantly better fitting models than the CDM models with less parameters.

For each of the CDM and ULDM models we perform MLEs assuming uniform priors on all parameters.  We also analyze other prior cases to test which galaxies or models are affected.  We constrain any parameters that are constrained from physical arguments. For the DC14 model, we constrain the free parameter $V_\text{200}$ such that $V_\text{200,min} \leq V_\text{200}$ where $V_\text{200,min}$ is found from the constraint that $\log_{10}\left(M_*/M_\text{halo}\right) < -1.3$.  For the ULDM models in which the soliton and halo profiles are matched, we take $\alpha$ for the single flavor model and $\alpha_1$ as well as $\alpha_2$ for the double flavor model to be fixed from Eq. (\ref{eq:matched_relation}).

For the analysis case in which we assume uniform priors on all parameters, we perform the fits constraining the free parameters as $1 \leq c_\text{200} \leq 100$, $1 \leq V_{200}/\left[\text{km} \, \text{s}^{-1}\right] \leq 1000$, $0.01 \leq \tilde{\Upsilon}_\text{d} \leq 5$ and $0.01 \leq \tilde{\Upsilon}_\text{b} \leq 5$, all of which have been taken in previous studies.  For the Einasto model, we take $\alpha$ to be unconstrained.  We find that we get similar results when taking the constraints $5 \times 10^{-3} \leq \alpha \leq 5$.  However, we leave the value of $\alpha$ unconstrained in order to allow more values for the ULDM models in which the soliton is matched to the halo. For the ULDM models, we take $10^{4.5} \leq M_\text{sol}/\left[M_\odot\right] \leq 10^{12}$.  When assuming the particle mass is free to vary in the fitting procedure, we take the particle masses within the range $10^{-3}\,m_{22} \leq m_i \leq 10^{3}\, m_{22}$.  We take the same range for one of the particle masses for the case in which the particle masses are treated as fixed parameters.  We fix the other particle mass to $m = 10^{1.5} \, m_{22}$, as we find that approximately around this particle mass, we obtain the best results.  After scanning this range, we find that only a small subset of masses produce reasonable fits for the ULDM models in which the soliton and halo are matched.  Therefore, for these models, we perform a more detailed scan in the mass range $m_{22} \leq m_2 \leq 10^2 \, m_{22}$, while fixing the other mass to be $m = 10^{1.5} \, m_{22}$.  We include a summary of all parameter ranges for all models tested in Table \ref{tab:params_sum}.

\renewcommand{\arraystretch}{2.5}
\begin{table}
\centering
\begin{tabular}{|c|c|c|c|c|c|c|c|c|c|c|c|c|}
\hline
& Burkert & DC14 & Einasto & NFW & SS(1) & SM(1) & DS(1) & DM(1) & SS(2) & SM(2) & DS(2) & DM(2) \\
\hhline{=============}
$c_{200,1}$ & \multicolumn{4}{c|}{$\left[1,100\right]$} & \multicolumn{4}{c|}{$\left[1,100\right]$} & \multicolumn{4}{c|}{$\left[1,100\right]$}\\
\hhline{-------------}
$c_{200,2}$ & \multicolumn{4}{c|}{-} & \multicolumn{2}{c|}{-} & \multicolumn{2}{c|}{$\left[1,100\right]$} & \multicolumn{2}{c|}{-} & \multicolumn{2}{c|}{$\left[1,100\right]$}\\
\hhline{-------------}
\Centerstack{$V_{200,1}$ \\ $\left[\text{km}\,\text{s}^{-1}\right]$} & $\left[1,1000\right]$ & $\left[V_{\text{m}},1000\right]$ & \multicolumn{2}{c|}{$\left[1,1000\right]$} & \multicolumn{4}{c|}{$\left[1,1000\right]$} & \multicolumn{4}{c|}{$\left[1,1000\right]$}\\
\hhline{-------------}
\Centerstack{$V_{200,2}$ \\ $\left[\text{km}\,\text{s}^{-1}\right]$} & \multicolumn{4}{c|}{-} & \multicolumn{2}{c|}{-} & \multicolumn{2}{c|}{$\left[1,1000\right]$} & \multicolumn{2}{c|}{-} & \multicolumn{2}{c|}{$\left[1,1000\right]$} \\
\hhline{-------------}
$\tilde{\Upsilon}_d$ & \multicolumn{4}{c|}{}  & \multicolumn{4}{c|}{}  & \multicolumn{4}{c|}{}\\
\hhline{-}
$\tilde{\Upsilon}_b$ & \multicolumn{4}{c|}{\multirow{-2}{*}{$\left[0.01,5\right]$}} & \multicolumn{4}{c|}{\multirow{-2}{*}{$\left[0.01,5\right]$}} & \multicolumn{4}{c|}{\multirow{-2}{*}{$\left[0.01,5\right]$}}\\
\hhline{-------------}
$\alpha_1$ & \multicolumn{2}{c|}{} & $\left(-\infty,\infty\right)$ & - & $\left(-\infty,\infty\right)$ & $\alpha_\text{M}$ & \multirow{2}{*}{} & \multirow{2}{*}{} & $\left(-\infty,\infty\right)$ & $\alpha_\text{M}$ & \multirow{2}{*}{} & \multirow{2}{*}{}\\
\hhline{-~~----~~--}
$\alpha_2$ & \multicolumn{2}{c|}{\multirow{-2}{*}{-}} & \multicolumn{2}{c|}{-} & \multicolumn{2}{c|}{-} & \multirow{-2}{*}{$\left(-\infty,\infty\right)$} & \multirow{-2}{*}{$\alpha_\text{M}$} & \multicolumn{2}{c|}{-}& \multirow{-2}{*}{$\left(-\infty,\infty\right)$} & \multirow{-2}{*}{$\alpha_\text{M}$}\\
\hhline{-------------}
\Centerstack{$m_1$ \\ $\left[m_{22}\right]$} & \multicolumn{4}{c|}{} & \multicolumn{4}{c|}{$\left[10^{-3},10^3\right]$} & $\left[10^{-3},10^3\right]$ & $\left[1,10^2\right]$ & \multicolumn{2}{c|}{$10^{1.5}$}\\
\hhline{-~~~~--------}
\Centerstack{$m_2$ \\ $\left[m_{22}\right]$} & \multicolumn{4}{c|}{\multirow{-2}{*}{-}} & \multicolumn{2}{c|}{-} & \multicolumn{2}{c|}{$\left[10^{-3},10^3\right]$} & \multicolumn{2}{c|}{-} & $\left[10^{-3},10^3\right]$ & $\left[1,10^2\right]$\\
\hhline{-------------}
\Centerstack{$M_\text{sol,1}$ \\ $\left[M_\odot\right]$} & \multicolumn{4}{c|}{} & \multicolumn{4}{c|}{$\left[10^{4.5},10^{12}\right]$} & \multicolumn{4}{c|}{$\left[10^{4.5},10^{12}\right]$}\\
\hhline{-~~~~--------}
\Centerstack{$M_\text{sol,2}$ \\ $\left[M_\odot\right]$} & \multicolumn{4}{c|}{\multirow{-2}{*}{-}} & \multicolumn{2}{c|}{-} & \multicolumn{2}{c|}{$\left[10^{4.5},10^{12}\right]$}& \multicolumn{2}{c|}{-} & \multicolumn{2}{c|}{$\left[10^{4.5},10^{12}\right]$}\\
\hhline{-------------}
\end{tabular}
\caption{Parameter ranges for each of the models tested.  Column 1 holds all of the necessary parameters, columns 2-5 hold all of the CDM models tested, and columns 6-13 hold all of the ULDM models tested.  For the ULDM models: SS corresponds to the single flavor and DS to the double flavor, summed models; SM corresponds to the single flavor, and DM to the double flavor, matched models; models with (1) correspond to the analysis in which the particle masses are free to vary in the fitting procedure; models with (2) correspond to the analysis in which the particle masses are fixed.  For the DC14 model, $V_{m}$ corresponds to the minimum virial velocity found from $\log_{10}\left(M_*/M_\text{halo}\right) < -1.3$.  For the ULDM matched models, $\alpha_\text{M}$ corresponds to the value of alpha fixed from Eq. (\ref{eq:matched_relation}).  For the ULDM models with (2), the values for $m_1$ and $m_2$ are fixed and the ranges quoted are scanned over during the fitting procedure.}
\label{tab:params_sum}
\end{table}

For all analyses, we check that the CMR \cite{Dutton:2014xda,Wang:2019ftp} is reproduced, that the resulting stellar and halo masses fit the AMR \cite{2013,2012}, that the baryonic mass and maximum circular velocity fit the BTFR \cite{2015}, that the distribution of mass-to-light ratios is consistent with stellar synthesis models \cite{2014}, and that the gravitational acceleration due to baryons and that due to DM fit the gravitational RAR of \cite{2016}. Each of these are described in more detail in App. \ref{sec:app_relations}. For the ULDM models, we also check that the SH relation (Eq. (\ref{eq:core_halo_relation})) is reproduced. 

\section{Comparison with previous studies} \label{sec:discussion}
We now discuss how our analysis compares to previous studies.  First, we extend the results of \cite{Bar:2018acw,Bar:2019bqz,10.1093/mnras/stw1256,Marsh:2015wka,10.1093/mnras/stx1941,Schive:2014dra,Safarzadeh:2019sre,2021ApJ...913...25C,Bar:2021kti,Dalal:2022rmp} by scanning over possible particles masses for double flavor ULDM models.  While the authors of \cite{Bar:2021kti} discuss the possible constraints on multiple flavored models, they do not perform a systematic scan over particle masses for multiple flavored models.  We therefore further their study by analyzing double flavor models and performing fits for all particle masses scanned.  We extend the results of these analyses by fitting ninety three galaxies in the SPARC catalog to single and double flavor ULDM models.  These ninety three galaxies have inclinations greater than $30^o$, have a measured value for the maximum circular velocity, $V_f$, in the SPARC catalog data, and have a quality flag that is not equal to three.  A quality flag of three corresponds to galaxies with either of the three:  major asymmetries; strong non-circular motions; or offests between HI measurements and stellar distributions \cite{Lelli:2016zqa}.  This generally means that the quoted measurements for the circular velocity may be unreliable.  These galaxies also have a total number of circular velocity measurements greater than eleven (for galaxies without bulge components) or twelve (for galaxies with bulge components), which is the number of parameters in the double flavor, summed model.

The authors of \cite{Broadhurst:2018fei} do consider multiple flavored models.  However, for most of the galaxies analyzed, they perform fits assuming that each galactic structure is composed of a single species of ULDM, while each galactic structure could potentially be composed of a different ULDM species.  They do, however, consider the MW to contain two solitonic structures composed of different ULDM species.  We extend the results of this analysis by fitting more galaxies in the SPARC catalog to double flavor ULDM models, in which each galaxy is assumed to be composed of two solitonic structures made up of different ULDM species.  We also, in addition to treating each mass as a free parameter in the fitting procedure, scan over possible particle masses for each galaxy.  In the next section, we further discuss the implications of treating each galactic structure as being composed of two species of ULDM. We note that our results may have been significantly different if we were to assume that each galaxy could be composed of either a single species or multiple species of ULDM and we leave an analysis of this sort for future work.

As opposed to some of the publications cited in this section, we do not assume that the SH relation is satisfied.  Rather we check that this relation is satisfied after performing fits.  Our reasoning is that the SH relation may be too restrictive, especially for some of the particle masses analyzed as well as for the double flavor models.  We also choose to model the ULDM halo profile with the Einasto profile rather than the NFW or Burkert models.  This is due, in part, to the fact that we find the Burkert and Einasto profiles to be the best performing CDM profiles analyzed in regards to the SPARC galaxies considered.  The Einasto profile has also been shown to produce good fits to simulated halos of CDM over a large range of masses \cite{Wang:2019ftp}.  

We extend the results of the studies cited in this section by analyzing both the reduced chi-square and BIC statistic of the resulting fits.  While the reduced chi-square can be utilized to compare the ULDM and CDM models, the BIC statistic has a more conservative penalization of models with more parameters, and therefore penalizes the ULDM models (especially the double flavor models).  We also extend previous studies by showing how the total sums of the reduced chi-square values over all galaxies analyzed depend on the ULDM masses scanned.  Finally, we show the resulting differences of the BIC statistics between the Einasto and ULDM models for the best fit particle masses found.

\section{Results}\label{sec:results}
Here, we show some of the results for the ULDM models while more results for the ULDM and CDM models can be found in App. \ref{sec:app_results}.  We show the results for ninety three galaxies in the SPARC catalog that have inclinations greater than $30^o$, have a measured value for the maximum circular velocity, $V_f$, in the SPARC catalog data, and have a quality flag that is not equal to three.  These galaxies also have a total number of circular velocity measurements greater than eleven (for galaxies without bulge components) or twelve (for galaxies with bulge components), which is the number of parameters in the double flavor, summed model.

We start with the comparison of the ULDM models to the CDM models.  We perform MLEs for the ULDM case in which particle mass is treated as a free parameter (ULDM models with (1) in Table \ref{tab:params_sum}) and show results for the ULDM models using the Einasto profile as the halo profile.  We compare this model to the Einasto only model as we find this and the Burkert model to perform better than the other CDM models analyzed (see App. \ref{sec:app_CDM_main_results} for results).  We choose to compare the ULDM models to the Einasto model due to the theoretical justification from \cite{Wang:2019ftp}. 

\begin{figure}
\centering
\makebox[0pt]{
\includegraphics[width=0.73\paperwidth]{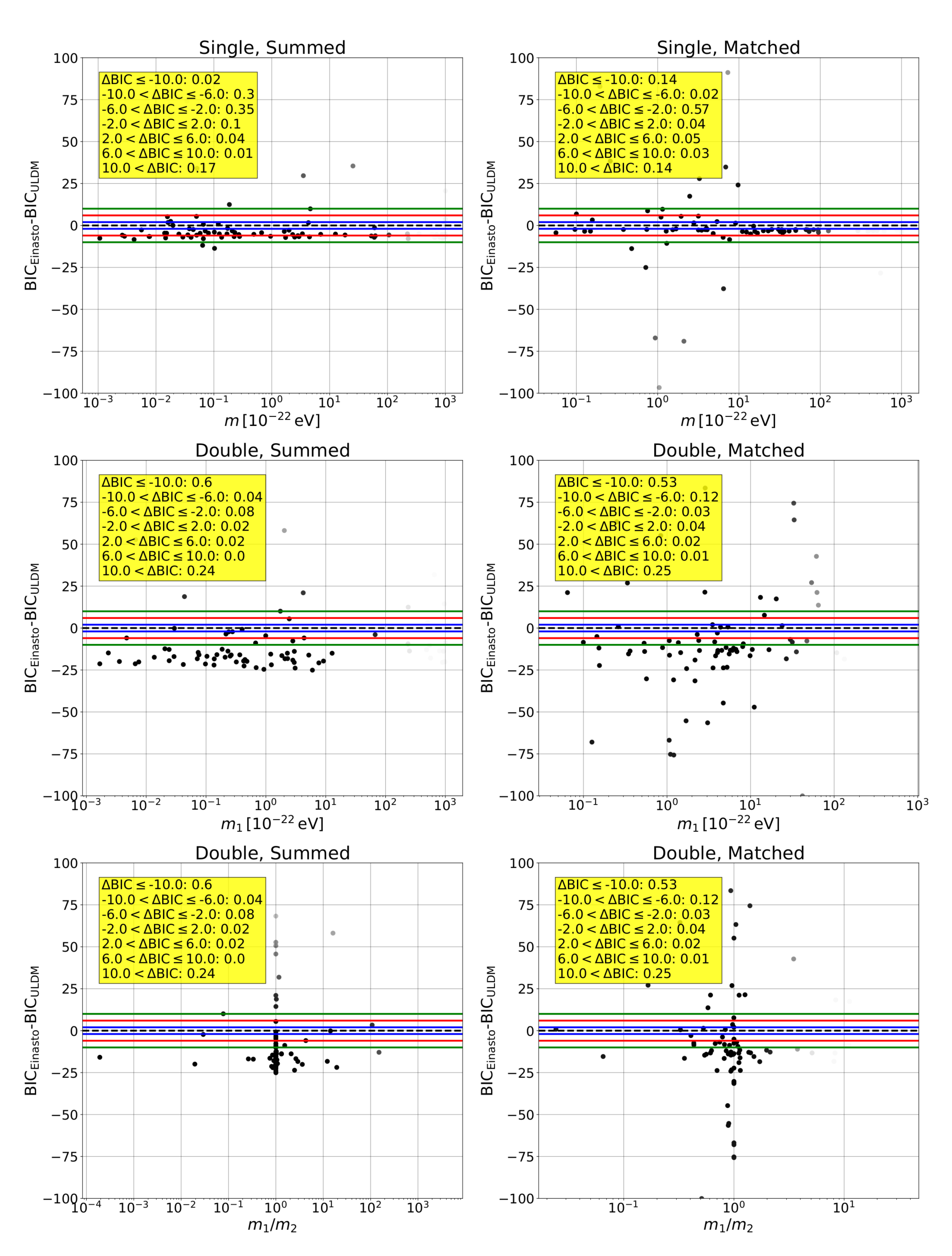}
}
\caption{\textbf{Particle masses free - Analysis (1) in Table \ref{tab:params_sum}: }Difference in the BIC statistics for Einasto and ULDM assuming the model:  single flavor, profiles summed (SS(1)) (top left); single flavor, profiles matched (SM(1)) (top right); double flavor, profiles summed (DS(1)) (middle and bottom left); double flavor, profiles matched (DM(1)) (middle and bottom right).  The middle panel corresponds to $m_1$, and the bottom panel to $m_1/m_2$. Black points correspond to each of the galaxies analyzed, the black dashed line corresponds to $\Delta$BIC$=0$, blue lines correspond to $\left|\Delta\text{BIC}\right|=2$, red lines to $\left|\Delta\text{BIC}\right|=6$, and green lines to $\left|\Delta\text{BIC}\right|=10$. Inset is the fraction of galaxies that fall within a given range for $\Delta \text{BIC}$ where the Einasto model is taken first in the difference.  The points are shaded corresponding to the approximate probability density, with darker points corresponding to denser regions.}
\label{fig:psi_mfree_BICvsm}
\end{figure}

Fig. \ref{fig:psi_mfree_BICvsm}~ shows the difference in the BIC statistic between the Einasto model and each of the ULDM models ($\Delta \text{BIC} = \text{BIC}_\text{Einasto} - \text{BIC}_\text{ULDM}$) vs. the particle mass.  We also show the lines of $\Delta\text{BIC}=0$, $\left|\Delta\text{BIC}\right| = 2$, $\left|\Delta\text{BIC}\right|=6$, and $\left|\Delta\text{BIC} \right|=10$ as the black dashed, blue, red, and green lines.  The fraction of galaxies that fall within a particular range for $\Delta \text{BIC}$ is shown in the inset, where the Einasto model is always taken first in the difference.  The points are shaded corresponding to the approximate probability density, with darker points corresponding to denser regions.

The top left panel of Fig. \ref{fig:psi_mfree_BICvsm}~ shows $\Delta \text{BIC}$ for the single flavored ULDM model in which the ULDM soliton and halo are summed togther (SS(1) model).  The largest fraction of galaxies $(35\%)$ shows a mild preference for the Einasto model, while the next largest fraction $(30\%)$ falls in the range of strong evidence in favor of the Einasto model.  Therefore, well over half of the galaxies analyzed show a preference for the Einasto model.  The next largest fraction $(17\%)$ falls in the range of decisive evidence for the SS(1) model, with the next largest fraction $(10\%)$ falling in the range of no preference for either model.  This suggests that the Einasto model is, in general, a better fitting model than the SS(1) model when taking into account the penalization of more model parameters.  

The middle and bottom panels (left column) of  Fig. \ref{fig:psi_mfree_BICvsm}~ show the double flavor model for which the soliton and halo are summed (DS(1) model), with the middle panel corresponding to $m_1$, and the bottom panel corresponding to $m_1/m_2$.  For this model, we obtain a large fraction $(60\%)$ of the galaxies showing decisive evidence for the Einasto model, while a total of $(72\%)$ of galaxies show some preference to decisive evidence for the Einasto model.  A little less than a quarter of the galaxies $(24\%)$ show decisive evidence for the DS(1) model.  Both summed models, then, result in the most galaxies showing some preference to decisive evidence for the Einasto model, when more model parameters are penalized.

The right column of Fig. \ref{fig:psi_mfree_BICvsm}~ shows the single and double flavor models for which the soliton and halo are matched (SM(1) and DM(1) models).  The SM(1) model performs better in some respects and worse in others than its summed counterpart (SS(1)), however the DM(1) model performs better overall than its summed counterpart (DS(1)).  This brings into question how the matched models would perform compared to the summed models if the matching relation (Eq. (\ref{eq:matched_relation})) were relaxed, and we discuss this later.  

Comparing the SM(1) model to the Einasto model, the largest fraction of galaxies $(57\%)$ falls in the range of mild preference for the Einasto model, with the next largest fractions ($14\%$) being equal and with one showing decisive evidence for the SM(1) model and the other showing decisive evidence for the Einasto model.  Comparing the DM(1) model to the Einasto model, a little over half of the galaxies $(53\%)$ show decisive evidence for the Einasto model, and a quarter of the galaxies $(25\%)$ show decisive evidence for the DM(1) model.  The next largest fraction $(12\%)$ shows strong evidence for the Einasto model.  

It is interesting to point out here that even though we let the particle masses vary in the range $10^{-3} \, m_{22} \leq m \leq 10^3 \, m_{22}$, for the matched models almost all galaxy fits prefer particle masses within the bounds of $10^{-1} \, m_{22} \lesssim m \lesssim 10^2$.  We explore this range of masses in more detail when we fix the particle masses in the fitting procedure.  Also, for the double flavored models, almost all galaxy fits prefer particle mass ratios in the range $10^{-2} \lesssim m_1/m_2 \lesssim 10$, with most galaxies showing a preference for approximately equal particle masses.  In our fitting procedure for the double flavor models, we take the initial guess for both of the soliton and particle masses to be equal.  The fact that the best fit parameters for both the particle masses happen to be approximately equal for many galaxies suggest that the choice of particle mass has little effect on the maximun likelihood estimates.  This is a reasonable suggestion, as the presence of the soliton will effect only the innermost regions, on the order of a kpc or less, where many galaxies have less data points.  We discuss this further when we discuss the error estimates for the best fit parameters.

\begin{figure}
\centering
\makebox[0pt]{
\includegraphics[width=0.75\paperwidth]{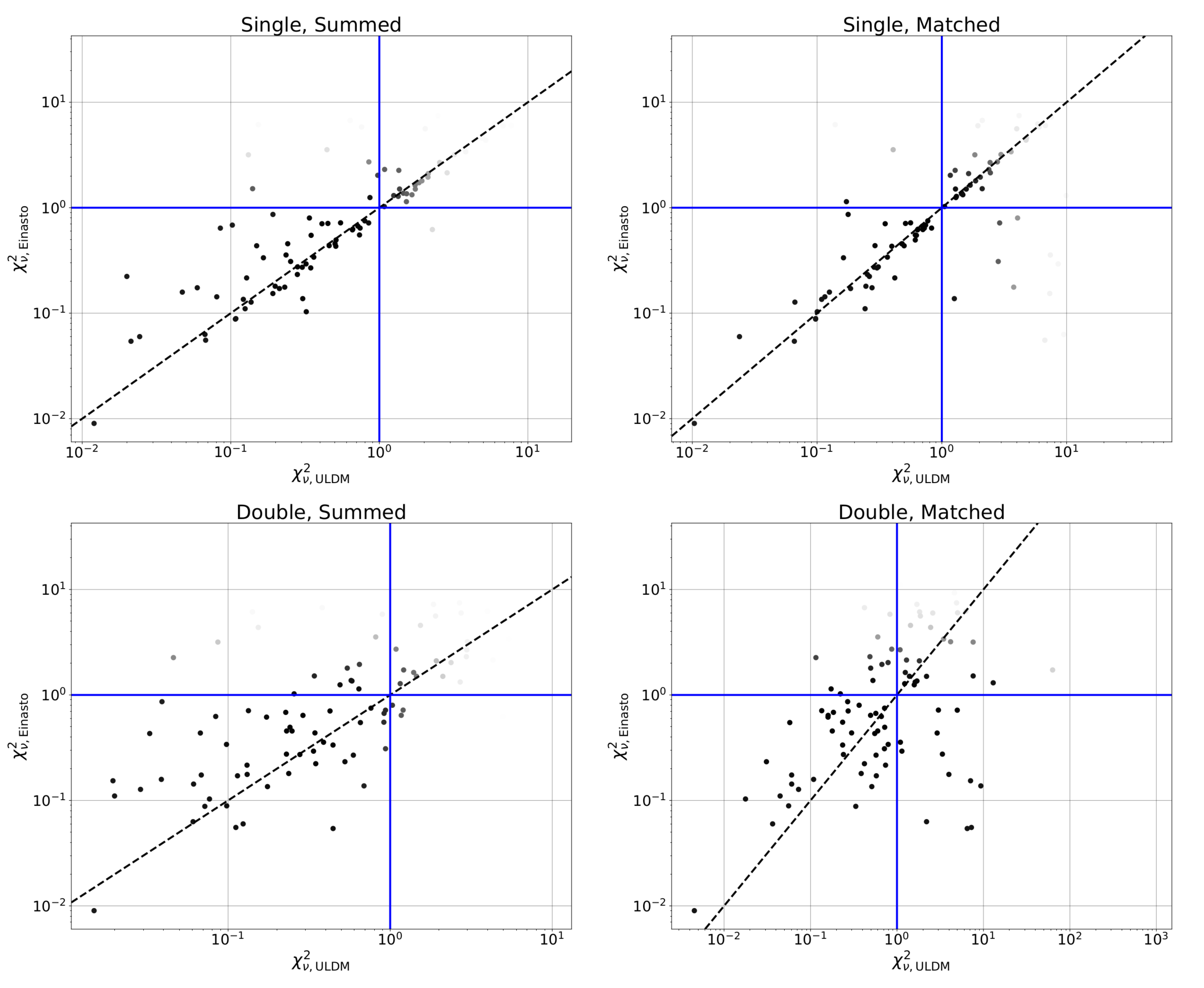}
}
\caption{\textbf{Particle masses free - Analysis (1) in Table \ref{tab:params_sum}: }Reduced chi-square for the Einasto model vs. reduced chi-square for the ULDM models:  SS(1) (top left); SM(1) (top right); DS(1) (bottom left); DM(1) (bottom right).  The black dashed lines correspond to $\chi^2_{\nu,\text{Einasto}} = \chi^2_{\nu,\text{ULDM}}$ while the blue horizontal lines show where $\chi^2_{\nu,\text{Einasto}} = 1$ and the blue vertical lines show where $\chi^2_{\nu,\text{ULDM}} = 1$.  The points are shaded corresponding to the approximate probability density, with darker points corresponding to denser regions.}
\label{fig:psi_mfree_chivschi}
\end{figure}

We now turn to the reduced chi-square statistic which does not penalize more model parameters as much as the BIC statistic.  Fig. \ref{fig:psi_mfree_chivschi}~ shows the reduced chi-square for the Einasto model vs. the reduced chi-square for each of the ULDM models.  The top left panel shows the SS(1) model, which has a significant number of galaxies giving reduced chi-square values for the Einasto model closer to one.  The top right panel shows the SM(1) model, which has a tighter correlation with $\chi^2_{\nu,\text{Einasto}} = \chi^2_{\nu,\text{ULDM}}$ than its summed counterpart.  The bottom left panel shows the DS(1) model, while the bottom right panel shows the DM(1) model.  In both cases, many galaxies give reduced chi-square values for the Einasto model that are closer to one.  For the matched model, many galaxies give reduced chi-square values greater than one, with some being significantly greater than one.    

\begin{figure}
\centering
\makebox[0pt]{
\includegraphics[width=0.55\paperwidth]{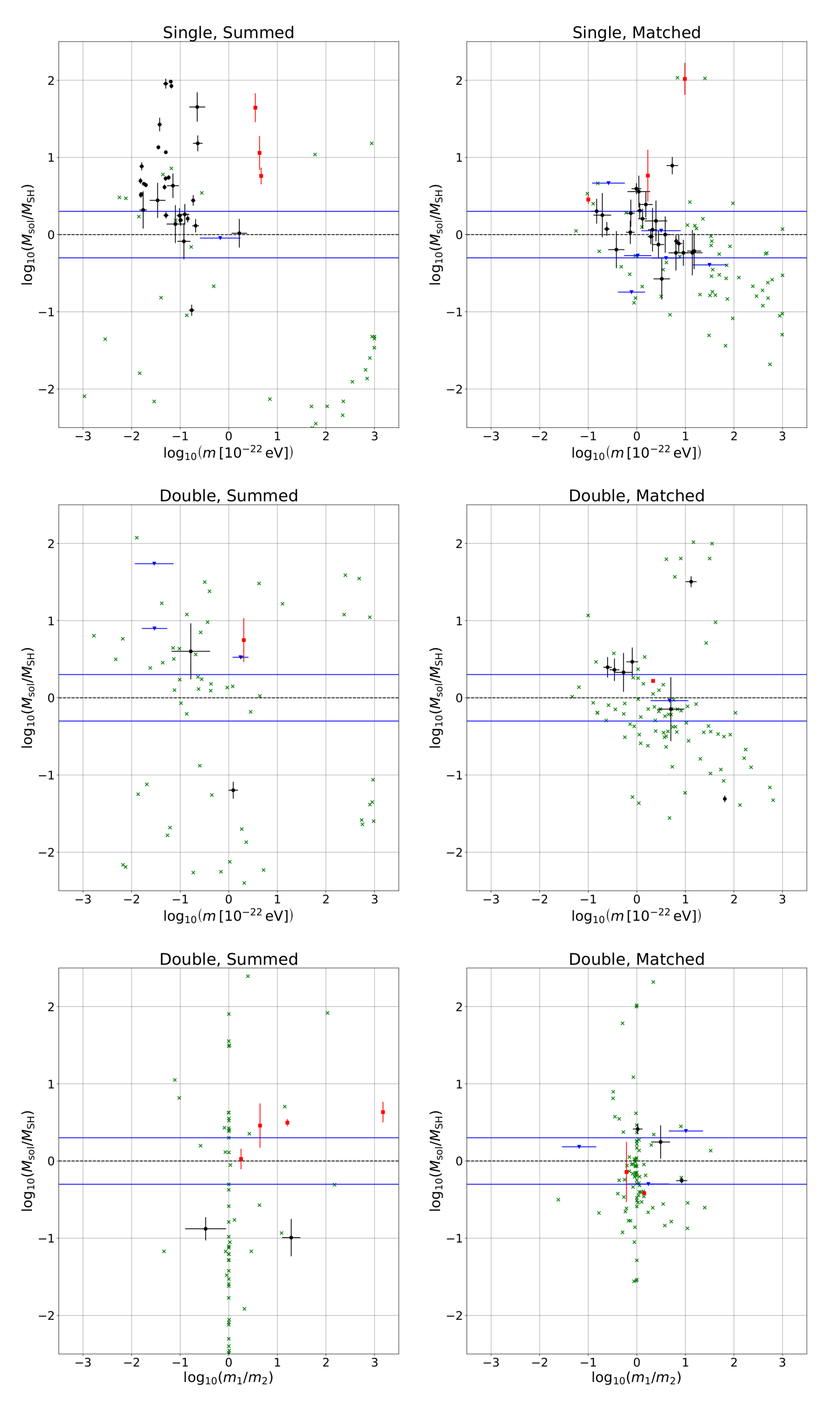}
}
\caption{\textbf{Particle masses free - Analysis (1) in Table \ref{tab:params_sum}: }$\log_{10}\left(M_\text{sol}/M_\text{sol,SH}\right)$ vs. $\log_{10} m$ where $M_\text{sol,SH}$ refers to the soliton mass assuming the SH relation given by Eq. (\ref{eq:core_halo_relation}) for the assumed models:  SS(1) (top left); SM(1) (top right); DS(1) (middle and bottom left) and DM(1) (middle and bottom right) with the middle row corresponding to $m_1$ and the bottom to $m_1/m_2$.  We plot each galaxy with a given marker depending on the error measurements for the particle and soliton masses.  The error calculated can either be:  (-) non-existent or larger than the best fit parameter; or (+) smaller than the best fit parameter.  Black points corresponds to galaxies with (+) for both the soliton and particle mass;  red squares correspond to galaxies with (+) for the soliton mass and (-) for the particle mass; blue triangles correspond to galaxies with (-) for the soliton mass and (+) for the particle mass; green x's correspond to galaxies with (-) for both the soliton and particle masses.}
\label{fig:psi_mfree_MsolvsMhalo}
\end{figure}

We now check the SH halo relation given by Eq. (\ref{eq:core_halo_relation}).  We compare the fit result soliton mass, $M_\text{sol}$, obtained in each ULDM model to the soliton mass as given by the SH relation, denoted as $M_\text{sol,SH}$.  For both the summed and matched models, the fit result soliton mass is the resulting best fit parameter.  Fig. \ref{fig:psi_mfree_MsolvsMhalo}~ shows the ratio (in log-10 space) between $M_\text{sol}$ and $M_\text{sol,SH}$ vs. particle mass.  The top row shows the single flavored model with the summed model (SS(1)) along the left column and the matched model (SM(1)) along the right column while the middle and bottom row correspond to the double flavored models (DS(1) on the left and DM(1) on the right).  The middle row corresponds to $m_1$, and the bottom row corresponds to $m_1/m_2$.  The blue lines allow for letting the SH relation differ by a factor of two.  The galaxies are marked with different markers depending on the error calculated in the fitting procedure, which we categorize as: (-) error measurements that are non-existent or larger than the best fit parameter; or (+) error measurements that are smaller than the best fit parameter.  Black points have (+) for both the soliton and particle masses; red squares have (+) for the soliton mass and (-) for the particle mass; blue triangles have (-) for the soliton mass and (+) for the particle mass; and green x's have (-) for both the soliton and particle masses.  

For the summed models (left panel of Fig. \ref{fig:psi_mfree_MsolvsMhalo}), there does not seem to be any trend of galaxies following the SH relation, while for the double flavor models, we obtain poor error measurements for many of the galaxies.  For the matched models (right panel), there is a tighter correlation with the SH relation, while the double flavor models again give many galaxies with poor error measurements.  Building on the suggestion above in regards to the fitting procedure choosing approximately equal particle masses for the double flavor model, the fact that many galaxies give poor error estimates futher confirms the suggestion that the MLEs are largely unaffected by changes in the soliton and particle masses. 

Treating the particle mass as free in the fitting procedure suggests the lack of any preference for a particular range of particle masses, within the range of masses searched, for the summed models.  However, for the DS(1) model, there is a significant preference for particle masses that are approximately equal.  For the matched models, on the other hand, there does seem to be a preference for a range of particle masses $m_{22} \lesssim m \lesssim 10^2$.  We, therefore, scan this mass range when we treat the particle masses as fixed parameters.  As in the summed model, there also seems to be a preference for approximately equal masses for the DM(1) model.

\begin{figure}
\centering
\makebox[0pt]{
\includegraphics[width=0.9\paperwidth]{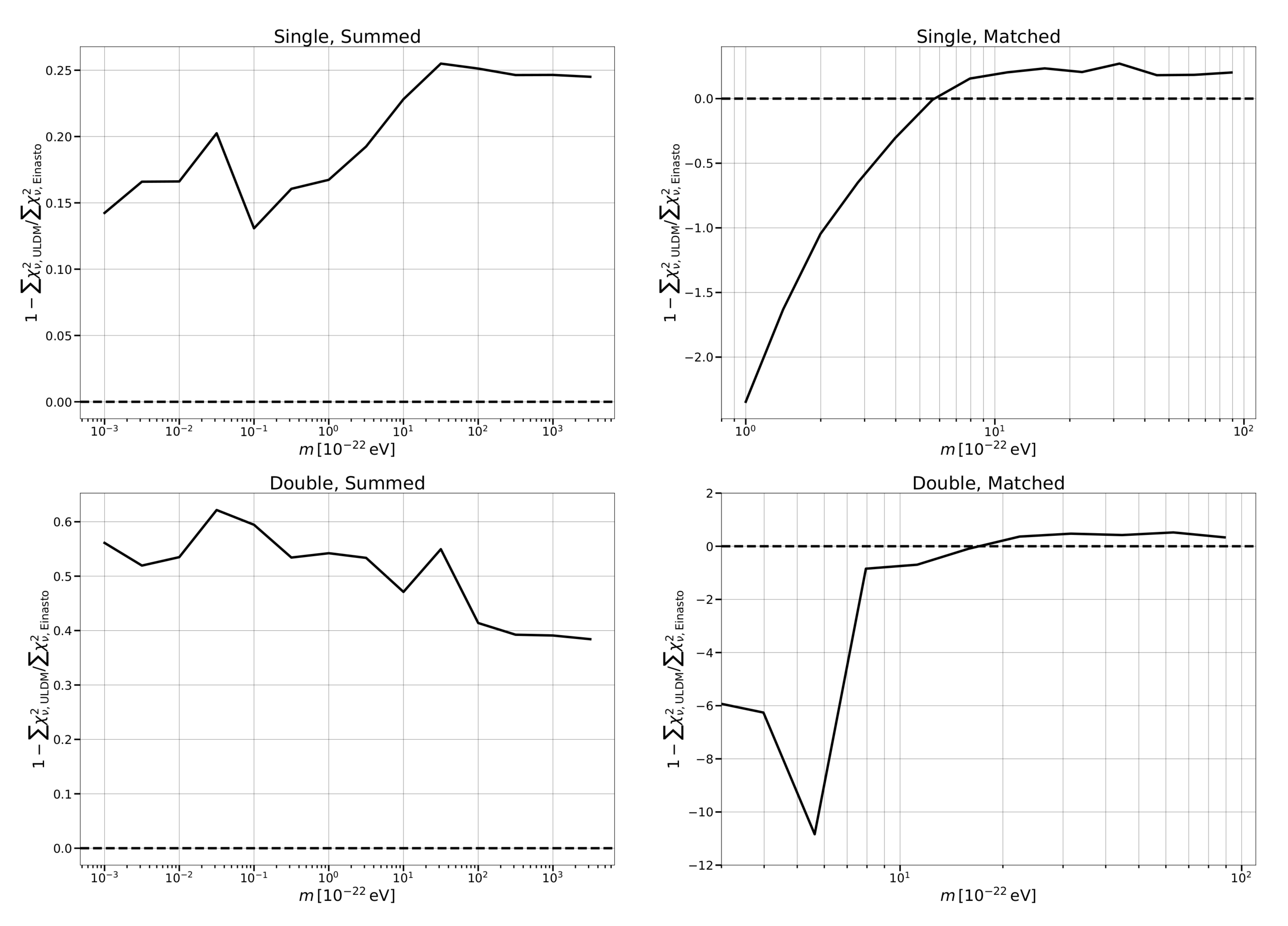}
}
\caption{\textbf{Particle masses fixed and scanned - Analysis (2) in Table \ref{tab:params_sum}: }$f(\chi^2_\nu) \equiv 1 - \sum\chi^2_{\nu,\text{ULDM}}/\sum\chi^2_{\nu,\text{Einasto}}$ vs. particle mass for the assumed models:  SS(2) (top left); SM(2) (top right); DS(2) (bottom left) and DM(2) (bottom right).}
\label{fig:psi_mfix_chisqvsm}
\end{figure}

We now discuss the ULDM models when the particle mass is fixed and scanned in the fitting procedure (ULDM models with (2) in Table \ref{tab:params_sum}).  Fig. \ref{fig:psi_mfix_chisqvsm}~ shows $f(\chi^2_\nu) \equiv 1 - \sum\chi^2_{\nu,\text{ULDM}}/\sum\chi^2_{\nu,\text{Einasto}}$ vs. fixed particle mass, where the sum is taken over all galaxies analyzed.  The single flavored models are shown on the top row with the summed model along the left column (SS(2)) and the matched model along the right column (SM(2)).  The double flavored models are shown on the bottom row (DS(2) on the left and DM(2) on the right).  

We find that the SS(2) model (top left panel of Fig. \ref{fig:psi_mfix_chisqvsm}) differs from the Einasto model by at most approximately $25\%$ and gives $f(\chi^2_\nu) > 0$ for all masses scanned.  The SM(2) model, on the other hand, differs more for masses $m \lesssim 3 \, m_{22}$.  All masses scanned in the range $6\, m_{22} \lesssim m \lesssim 10^2 \, m_{22}$ give $f(\chi^2_\nu) > 0$.  For both the SS(2) and SM(2) models, we find the best fit mass to be $m \approx 10^{1.5} \, m_{22}$.  We then fix one of the particle masses to this particular particle mass when analyzing the double flavor models (DS(2) and DM(2)).

The bottom row of Fig. \ref{fig:psi_mfix_chisqvsm} shows the double flavor models for which one of the particle masses is fixed to $m_1 = 10^{1.5} \, m_{22}$, and the other is scanned over a particular range (see Table \ref{tab:params_sum} for ranges).  The bottom left panel shows the DS(2) model which gives $f(\chi^2_\nu) > 0$ for all masses scanned.  This model differs from the Einasto model by at most approximately $60\%$ and by at least approximately $40\%$.  The DM(2) model, on the other hand, differs from the Einasto model significantly for masses $m_2 \lesssim 10 \, m_{22}$.  The DM(2) model gives $f(\chi^2_\nu) > 0$ for masses $11 \, m_{22} \lesssim m_2 \lesssim 10^2 \, m_{22}$, and gives the best results for masses $m_1 = 10^{1.5}\,m_{22}$ and $m_2 = 10^{1.8}\,m_{22}$.  Later, we show some results assuming these fixed masses.

\begin{figure}
\centering
\makebox[0pt]{
\includegraphics[width=0.7\paperwidth]{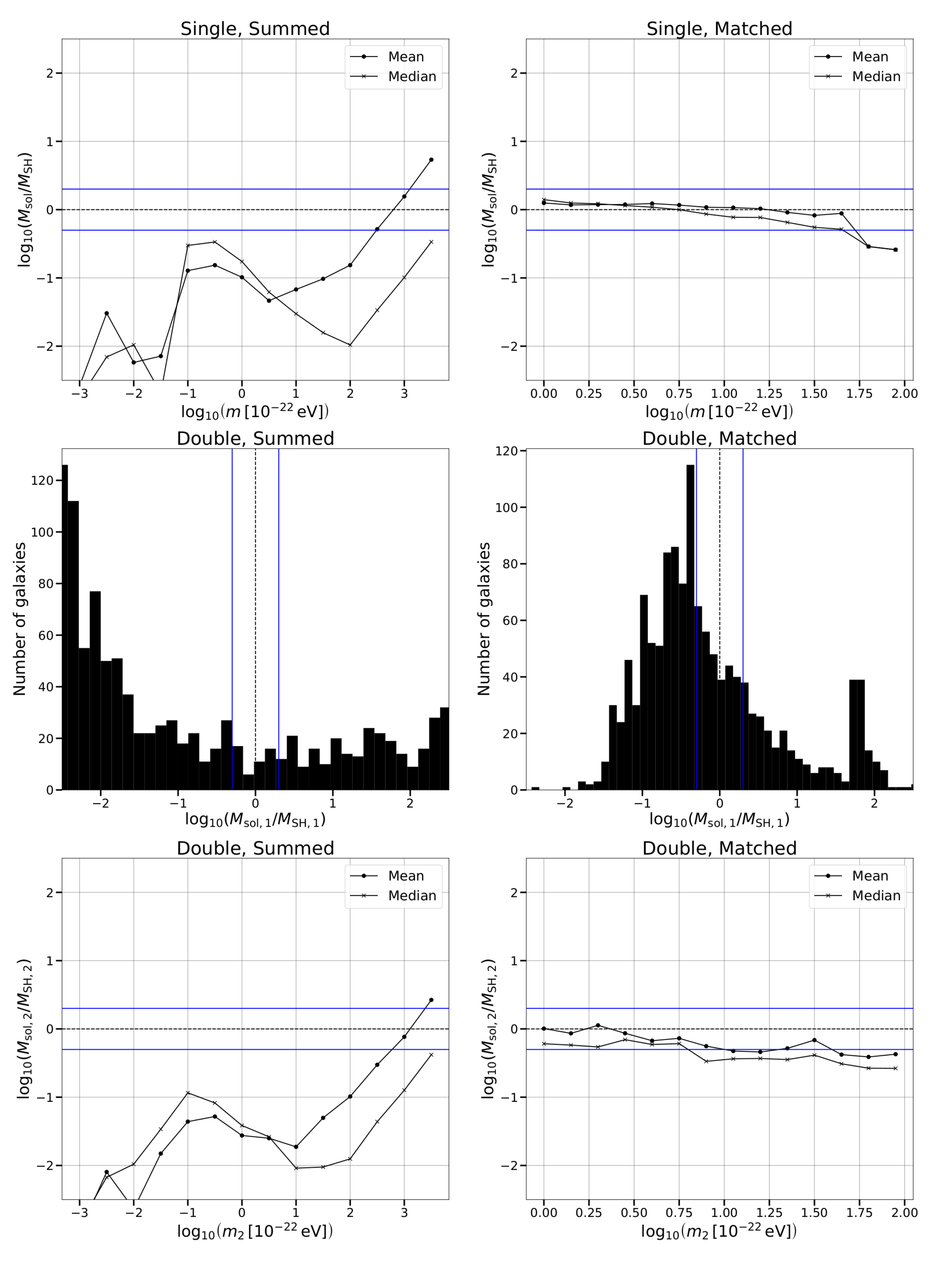}
}
\caption{\textbf{Particle masses fixed and scanned - Analysis (2) in Table \ref{tab:params_sum}: }$\log_{10}\left(M_\text{sol}/M_\text{sol,SH}\right)$ vs. $\log_{10}m$ (top and bottom panels) where $M_\text{sol,SH}$ refers to the soliton mass assuming the SH relation given by Eq. (\ref{eq:core_halo_relation}).  The top row corresponds to the SS(2) (left) and SM(2) (right) models.  The middle and bottom rows correspond to the DS(2) (left) and DM(2) (right) models.  The middle row corresponds to the particle mass fixed at $m = 10^{1.5} \, m_{22}$ and shows the histogram of $\log_{10}\left(M_\text{sol}/M_\text{sol,SH}\right)$ for this particular mass.  The bottom row corresponds to the particle mass that is varied in the double flavor models.  For the top and bottom rows, the black points(x's) correspond to the mean(median) value of $\log_{10}\left(M_\text{sol}/M_\text{sol,SH}\right)$.}
\label{fig:psi_mfix_MsolvsMhalo}
\end{figure}

The results discussed above are obtained with no dependence on the SH relation.  We now show how the results compare to this relation given by Eq. (\ref{eq:core_halo_relation}).  As in the analysis case for which the particle mass was free to vary in the fitting procedure, we compare the fit result soliton mass, $M_\text{sol}$, to the soliton mass as given by the SH relation, $M_\text{sol,SH}$. Fig. \ref{fig:psi_mfix_MsolvsMhalo}~ shows the ratio (in log-10 space) of $M_\text{sol}$ to $M_\text{sol,SH}$ vs. the particle mass for the top and bottom rows.  Black points(x's) correspond to the mean(median) vaue of $\log_{10}\left(M_\text{sol}/M_\text{sol,SH}\right)$.  The middle row shows the distribution of $\log_{10}\left(M_\text{sol}/M_\text{sol,SH}\right)$ for the fixed particle mass $m_1 = 10^{1.5}\, m_{22}$ for the double flavor model.  The number of samples corresponds to the number of galaxies analyzed for each possible $m_1$, $m_2$ pair, with $m_2$ varied along the same range as in the single flavored model.  The top row shows the single flavored models, with the SS(2) model on the left and SM(2) model on the right.  The SS(2) model gives both the mean and median outside of the SH relation range for almost all particle masses scanned, while the SM(2) model results in almost all masses scanned giving the mean and median values falling within the SH relation range.  We see the same sort of behavior for double flavor models.

\begin{figure}
\centering
\makebox[0pt]{
\includegraphics[width=0.7\paperwidth]{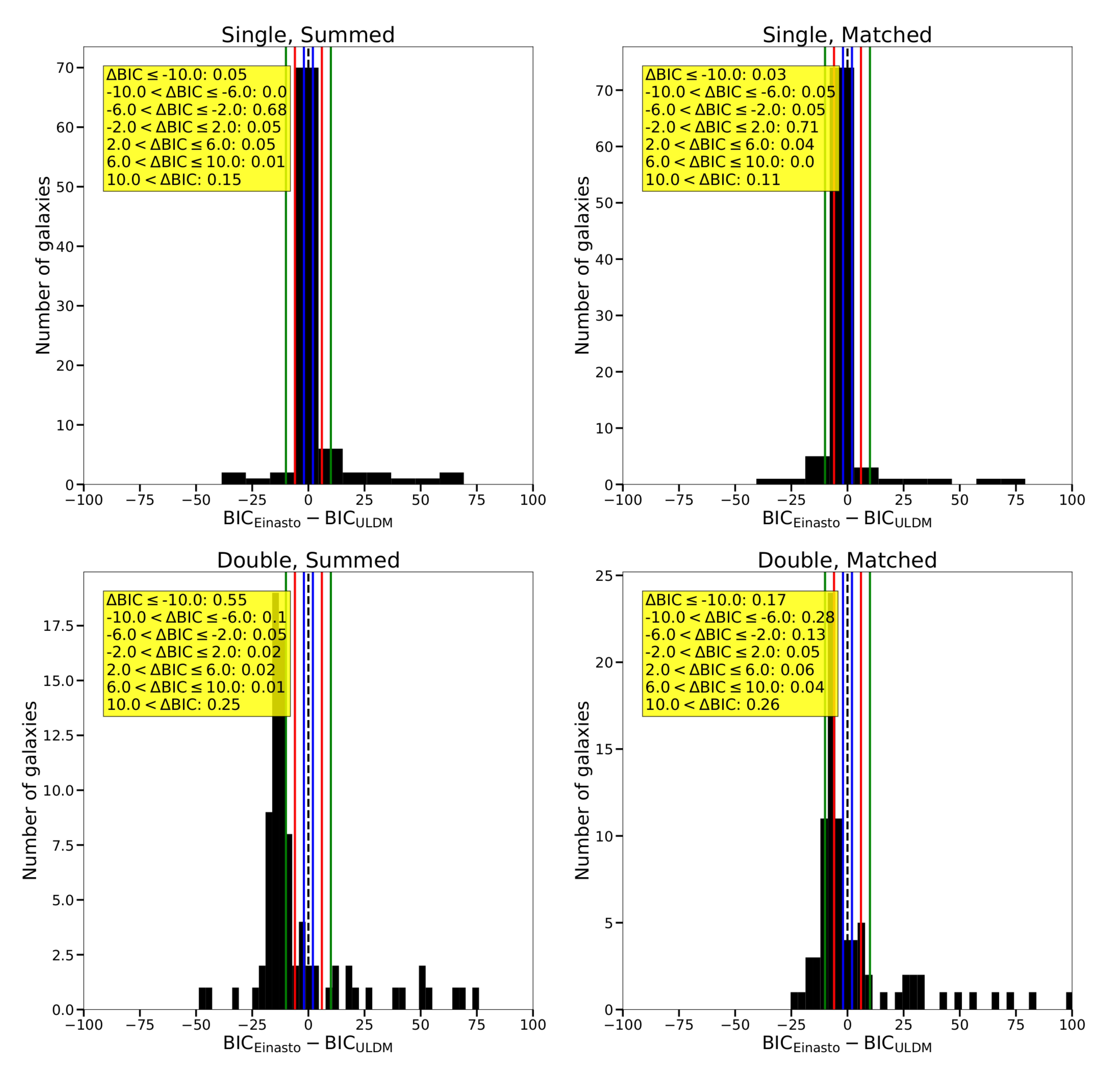}
}
\caption{\textbf{Particle masses fixed ($m_1 = 10^{1.5}\,m_{22}$, $m_2 = 10^{1.8}\, m_{22}$): }Difference in the BIC statistics for Einasto and ULDM assuming the model:  SS (top left); SM (top right); DS (bottom left); DM (bottom right).  The black dashed line corresponds to $\Delta$BIC$=0$, blue lines correspond to $\left|\Delta\text{BIC}\right|=2$, red lines to $\left|\Delta\text{BIC}\right|=6$, and green lines to $\left|\Delta\text{BIC}\right|=10$. Inset is the fraction of galaxies that fall within a given range for $\Delta \text{BIC}$ where the Einasto model is taken first in the difference.}
\label{fig:psi_mfix_BICvsBIC_ex}
\end{figure}

Finally, we fix the particle masses to $m_1 = 10^{1.5} \, m_{22}$ for all ULDM models and $m_2 = 10^{1.8}\,m_{22}$ for the double flavor models and vary all the rest of the parameters as in analysis (2) in Table \ref{tab:params_sum}.  Fig. \ref{fig:psi_mfix_BICvsBIC_ex}~ shows $\Delta \text{BIC} = \text{BIC}_\text{Einasto} - \text{BIC}_\text{ULDM}$ for the SS (top left), SM (top right), DS (bottom left), and DM (bottom right) models.  Comparing this to Fig. \ref{fig:psi_mfree_BICvsm}, one can see that all models besides the DS model perform better when the masses are fixed in this way rather than being allowed to vary in the fitting procedure, with the matched models performing siginificantly better.

\begin{figure}
\centering
\makebox[0pt]{
\includegraphics[width=0.8\paperwidth]{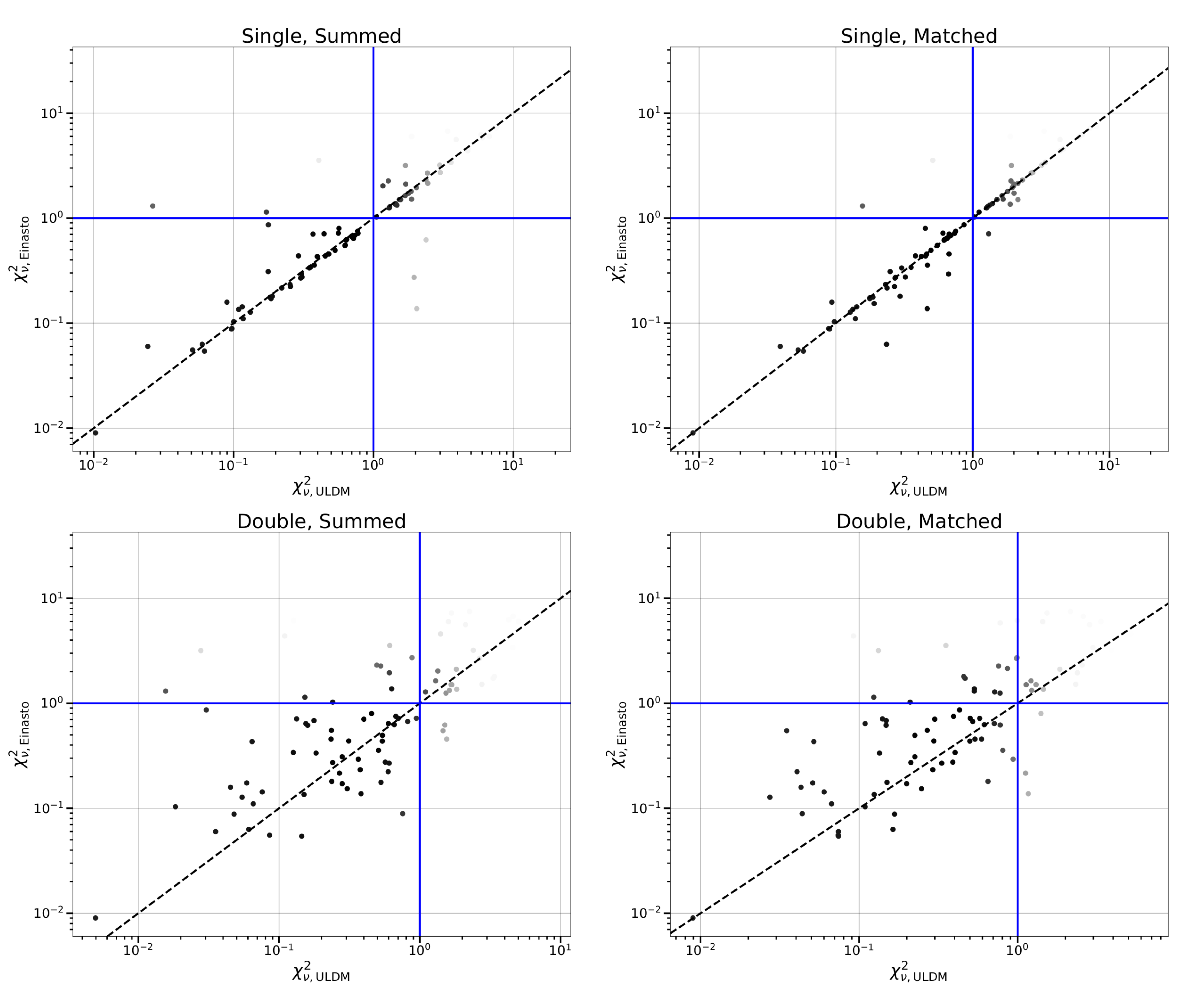}
}
\caption{\textbf{Particle masses fixed ($m_1 = 10^{1.5}\,m_{22}$, $m_2 = 10^{1.8}\, m_{22}$): }Reduced chi-square for the Einasto model vs. reduced chi-square for the ULDM models:  SS (top left); SM (top right); DS (bottom left); DM (bottom right).  The black dashed lines correspond to $\chi^2_{\nu,\text{Einasto}} = \chi^2_{\nu,\text{ULDM}}$ while the blue horizontal lines show where $\chi^2_{\nu,\text{Einasto}} = 1$ and the blue vertical lines show where $\chi^2_{\nu,\text{ULDM}} = 1$.  The points are shaded corresponding to the approximate probability density, with darker points corresponding to denser regions.}
\label{fig:psi_mfix_chivschi_ex}
\end{figure}

We can also compare the reduced chi-square results to those for which the particle mass is allowed to vary in the fitting procedure (Fig. \ref{fig:psi_mfree_chivschi}).  Fig. \ref{fig:psi_mfix_chivschi_ex}~ shows the reduced chi-square for the Einasto model vs. the reduced chi-square for the ULDM models, again fixing the particle masses to $m_1 = 10^{1.5} \, m_{22}$ and $m_2 = 10^{1.8}\, m_{22}$.  Both the SS (top left) and SM (top right) models have most galaxies tightly correlated with  $\chi^2_{\nu,\text{Einasto}} = \chi^2_{\nu,\text{ULDM}}$, with the matched model giving a handful of galaxies that result in a reduced chi-square closer to one than the Einsato model.  The DS (bottom left) model performs a bit better than the case in which the particle masses are allowed to vary, with more galaxies giving a reduced chi-square closer to one than the Einasto model.  The DM (bottom right) model performs significantly better than the case in which the particle masses are allowed to vary, with most galaxies that gave significantly large reduced chi-squares now giving reduced chi-squares closer to one.

\begin{figure}
\centering
\makebox[0pt]{
\includegraphics[width=0.6\paperwidth]{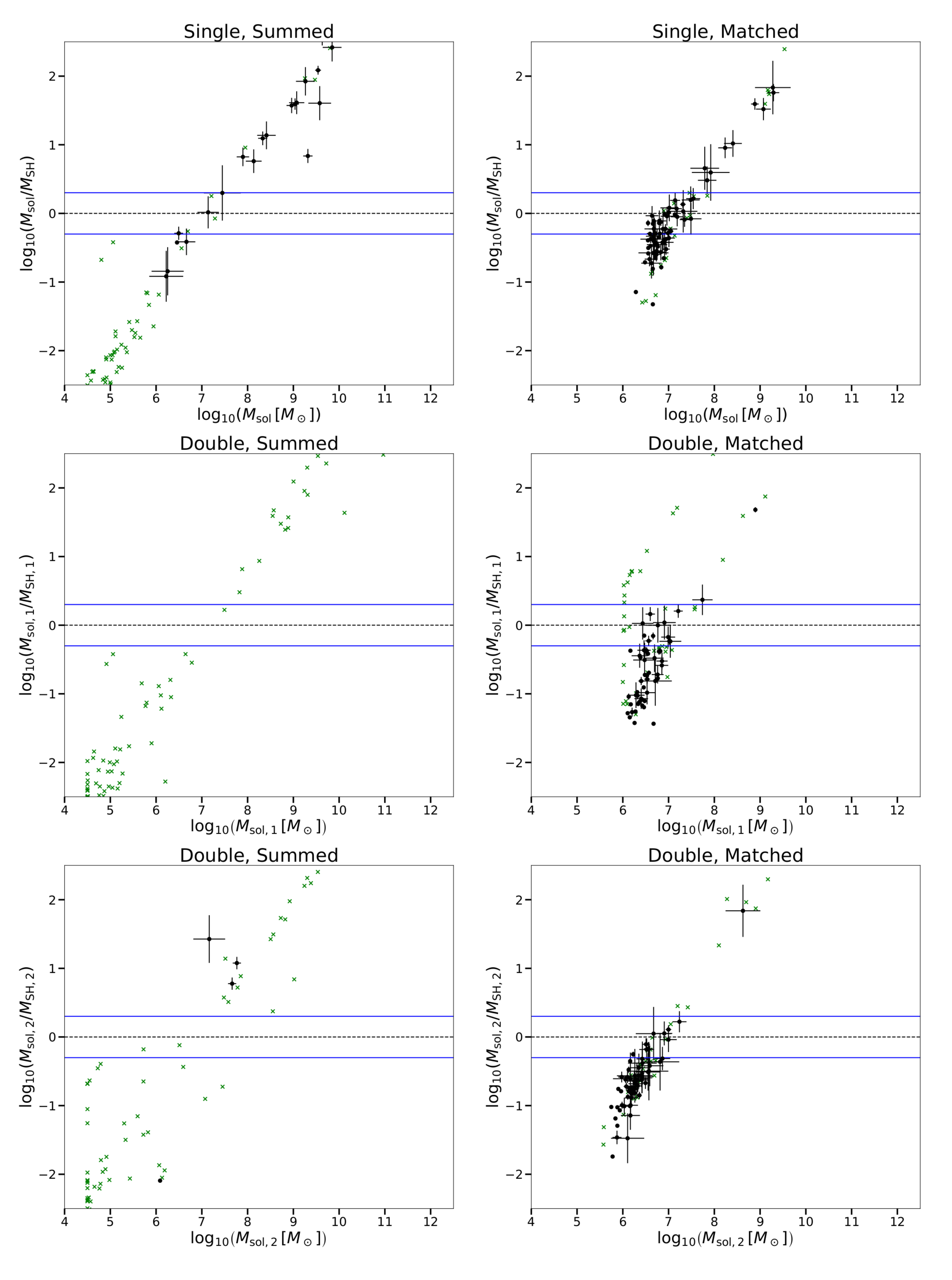}
}
\caption{\textbf{Particle masses fixed ($m_1 = 10^{1.5}\,m_{22}$, $m_2 = 10^{1.8}\, m_{22}$): }$M_\text{sol}/M_\text{sol,SH}$ where $M_\text{sol,SH}$ refers to the soliton mass assuming the soliton halo relation given by Eq. (\ref{eq:core_halo_relation}).  The top row corresponds to the SS (left) and SM (right) models.  The middle and bottom rows correspond to the DS (left) and DM (right) models.  The top and middle row corresponds to the particle mass fixed at $m = 10^{1.5} \, m_{22}$, and the bottom row to the particle mass fixed at $m = 10^{1.8}\,m_{22}$.  The points are marked in the same way as in Fig. \ref{fig:psi_mfree_MsolvsMhalo}.}
\label{fig:psi_mfix_MsolvsMhalo_ex}
\end{figure}

Finally, we show how the resulting soliton masses compare to the SH relation in Fig. \ref{fig:psi_mfix_MsolvsMhalo_ex} when the particle masses are fixed to $m_1 = 10^{1.5}\,m_{22}$ and $m_2 = 10^{1.8}\, m_{22}$.  These results can be compared to the case in which the particle mass is allowed to vary (Fig. \ref{fig:psi_mfree_MsolvsMhalo}) and the case in which the particle masses are fixed, but scanned in the fitting procedure (Fig. \ref{fig:psi_mfix_MsolvsMhalo}).  The points are marked in the same way as in Fig. \ref{fig:psi_mfree_MsolvsMhalo}.  When the particle masses are fixed in this way, we find that again the summed models have a larger variance with respect the SH relation compared to the matched model, while we again find many galaxies giving poor error measurements for the double flavor summed model.  On the other hand, the double flavor matched model gives significantly more galaxies with reasonable error measurements calculated in the fitting procedure (compared to the DM(1) analysis).  This suggests that when the particle masses are fixed in this way, the MLEs have a larger dependence on the soliton mass. 

We refer the reader to App. \ref{sec:app_results} for many more results.  We include the comparisons with other empirical relations (i.e. the CMR, BTFR, AMR, and gravitational RAR), statistical and parameter distributions for all models analyzed, and various rotation curves.  We also note that some galaxies exhibit a strong degeneracy in the best fit values for $\tilde{\Upsilon}_d$.  In this case, it is possible for the fitting routine to choose best fit values for $\tilde{\Upsilon}_d$ that are at the minimum or maximum values.

\section{Conclusion}\label{sec:conclusion}
We compared single and double flavor ultralight dark matter (ULDM) models of galactic dark matter to each other and to commonly used CDM models.  We fit these models to the measured galactic circular velocities of galaxies in the SPARC catalog, and compare models using the reduced chi-square and BIC statistics.  We analyzed cases for which the particle masses in the ULDM models are free to vary, and for which the particle masses are fixed in the fitting procedure.  For each of these analyses, we perform fits for ULDM models in which the soliton and halo are summed together; and for ULDM models in which the soliton and halo are matched.  

When the particle mass was free in the fitting procedure, we found that there is a negligible preference for any particular range of particle masses, within $10^{-25} \, \text{eV} \leq m \leq 10^{-19} \, \text{eV}$, when assuming the summed models.  For the matched models, however, we found that almost all galaxies prefer particles masses in the range  $10^{-21} \, \text{eV} \lesssim m \lesssim 10^{-20} \, \text{eV}$.  For both double flavor models (summed and matched) we found that most galaxies prefer approximately equal particle masses.  We also found that many galaxies tend to fall outside the range of the soliton-halo (SH) relation for all models analyzed.  For all analyses, the summed models gave much larger variances with respect to the SH relation than the matched models.

When the particle masses were fixed in the fitting procedure, we found that both single flavor models gave a maximum in $f(\chi^2_\nu) \equiv 1 - \sum\chi^2_{\nu,\text{ULDM}}/\sum\chi^2_{\nu,\text{Einasto}}$, where the sum is taken over all galaxies, for the particle mass $m=10^{1.5}\,m_{22}$.  The single flavor, summed models gave $f(\chi^2_\nu) > 0$ for all masses scanned, while the single flavor, matched model gave $f(\chi^2_\nu) > 0$ for all masses scanned in the range $6\, m_{22} \lesssim m \lesssim 10^2 \, m_{22}$.  For the double flavor models, we fixed one of the particle masses to the best fit particle mass $m_1 = 10^{1.5}\,m_{22}$.  The double flavor, summed model gave $f(\chi^2_\nu) > 0$ for all masses scanned, while the double flavor, matched model gave $f(\chi^2_\nu) > 0$ for masses $11 \, m_{22} \lesssim m_2 \lesssim 10^2 \, m_{22}$.  The double flavor, matched model gave the best fit results for masses $m_1 = 10^{1.5}\,m_{22}$ and $m_2 = 10^{1.8}\,m_{22}$.  As in the analysis case for which the particle masses were free to vary in the fitting procedure, many galaxies fell outside the range of the SH relation.  However, the mean and median value for all galaxies analyzed tended to follow the SH relation assuming the matched models for almost all masses scanned.

The results shown is this study were based on different assumptions that can be changed.  First, it is important to note that one can treat the point at which the soliton and halo are matched as a free parameter.   It is also possible to take into account the fact that some galaxies may be better fit by a single flavor, and some to a double flavored model, while each galaxy could have differing radii at which the soliton and halo are matched.  Finally, one can also fit each galaxy based on the fit parameters of the last galaxies, a problem that can be handled particularly well using reinforcement learning.  In this case, one may find a set of parameters that better fit more galaxies on average.  These possibilites result in a complex map of ULDM halos dependent on the abundance of the ULDM species and the collapse history of the ULDM halo.  We discuss these possibilites in a future study in which we utilize a reinforcement learning algorithm to take into account the complex map of possibilities and infer the possible ULDM abundances present today.

\section{Acknowledgments}
This material is based upon work supported by the U.S. Department of Energy (DOE), Office of Science, Office of Workforce Development for Teachers and Scientists, Office of Science Graduate Student Research (SCGSR) program.  The SCGSR program is administered by the Oak Ridge Institute for Science and Education (ORISE) for the DOE.  ORISE is managed by ORAU under contract number DE-SC0014664.  All opinions expressed in this paper are the authors' and do not necessarily reflect the policies and views of DOE, ORAU, or ORISE.  L.S. thanks Joshua Eby and Peter Suranyi for valuable discussions.  L.S. thanks Mike Sokoloff and Daniel Vieira for setting up computational resources to be used in the next installment of this study.

\bibliography{DM_halos}
\bibliographystyle{JHEP}

\clearpage
\begin{appendix}
\section{Relations}\label{sec:app_relations}
For all analyses, we check that the distribution of mass-to-light ratios is consistent with stellar synthesis models \cite{2014}, that the resulting stellar and halo masses fit the abundance matching relation (AMR) \cite{2013,2012}, that the baryonic mass and maximum circular velocity fit the baryonic Tully-Fisher relation (BTFR) \cite{2015}, that the gravitational acceleration due to baryons and that due to DM fit the gravitational radial acceleration relation (RAR) of \cite{2016}, and that the concentration mass relation (CMR) \cite{Dutton:2014xda,Wang:2019ftp} is reproduced.  For the ULDM models, we also check that the soliton-halo (SH) relation (Eq. (\ref{eq:core_halo_relation})) is reproduced.

The RAR of \cite{2016} is given by,
\begin{align}\label{eq:grar}
g_\text{tot}(r) = g_\text{bar}\left(1-e^{-\sqrt{g_\text{bar}(r)/g_{\dagger}}}\right)^{-1},
\end{align}
where $g_\text{tot}$ is the total gravitational acceleration and $g_\text{bar}$ is that due to baryons at a given radial distance $r$ and $g_\dagger$ was fit to be $g_{\dagger} = 1.2\times10^{-10}\,\text{m}/\text{s}^2$.  

The concentration mass relation of \cite{Dutton:2014xda} is given by,
\begin{align}\label{eq:conc_mass_relation_Du}
\log_{10}c_{200} = 0.905 - 0.101 \log_{10} \left(\frac{M_{200}}{10^{12} h^{-1} M_\odot}\right),
\end{align}
with a scatter of 0.11 dex.  The concentration mass relation of \cite{Wang:2019ftp} is given by,
\begin{align}\label{eq:conc_mass_relation_Wa}
c_{200} = \sum_{i=0}^5 c_i \ln \left(\frac{M_{200}}{h^{-1} M_\odot}\right),
\end{align}
where $c_i = \left[27.112, -0.381, -1.853 \times 10^{-3}, -4.141 \times 10^{-4}, -4.334 \times 10^{-6}, 3.208 \times 10^{-7}\right]$ for $i\in \left\{0,5\right\}$. 

The baryonic Tully-Fisher relation of \cite{2015} is given by,
\begin{align}\label{eq:BTFR}
\log_{10} \left(\frac{M_b}{M_\odot}\right) = s \log_{10} \left(\frac{V_f}{\text{km}/\text{s}}\right) + \log_{10} A,
\end{align}
where $M_b$ is the baryonic mass, $V_f$ is the maximum circular velocity, and the fit parameters were found to be $s = 3.71 \pm 0.08$ and $\log_{10} A = 2.27 \pm 0.18$.  

Finally, the abundance matching relation between stellar and DM masses is given by \cite{2013,2012},
\begin{align}\label{eq:AMR}
\frac{M_*}{M_{200}} = 2 N \left[\left(\frac{M_{200}}{M_1 M_\odot}\right)^{-\beta} + \left(\frac{M_{200}}{M_1 M_\odot}\right)^{-\gamma}\right]^{-1},
\end{align}
where $M_*$ is the total stellar mass, $M_{200}$ is the DM halo mass, $N = 0.0351$, $\beta = 1.376$, $\gamma = 0.608$, and $\log_{10}(M_1) = 11.59$.

\section{Results}\label{sec:app_results}
Here, we show many more results including the single and ULDM models \ref{sec:app_psi}, a comparison with previous studies \ref{sec:app_comp_study}, and results for the CDM models \ref{sec:app_CDM_main_results}.  

\subsection{ULDM}\label{sec:app_psi}
\begin{figure}
\centering
\makebox[0pt]{
\includegraphics[width=0.35\paperwidth]{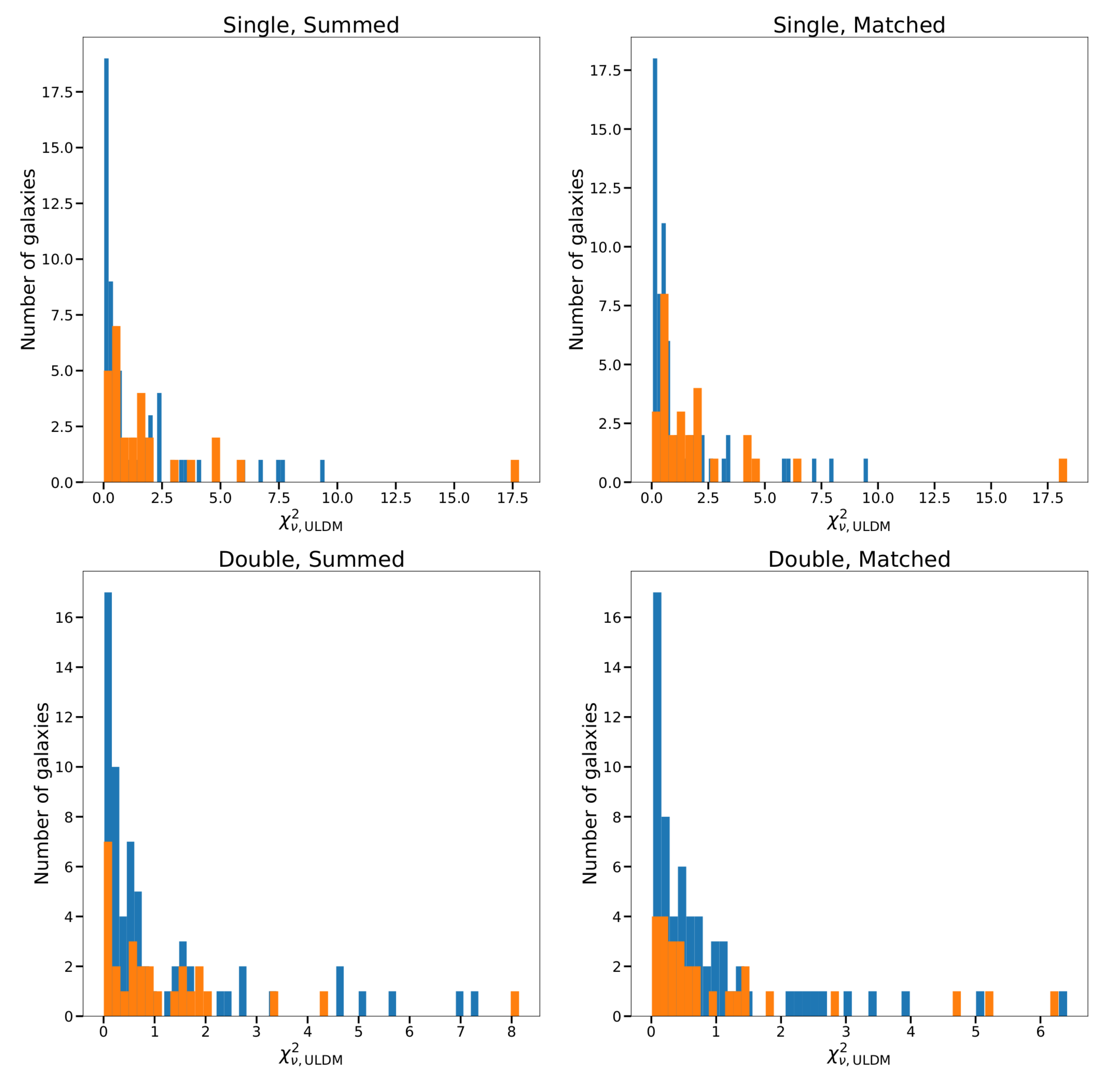}
}
\caption{\textbf{Particle masses fixed ($m_1 = 10^{1.5}\,m_{22}$, $m_2 = 10^{1.8}\, m_{22}$): }Reduced chi-square $\chi^2/\nu$ for the assumed models: single, summed (top, left); single, matched (top, right); double, summed (bottom, left); double, matched (bottom, right).}
\label{fig:psi_mfix_dist_ex}
\end{figure}

\begin{figure}
\centering
\makebox[0pt]{
\includegraphics[width=0.55\paperwidth]{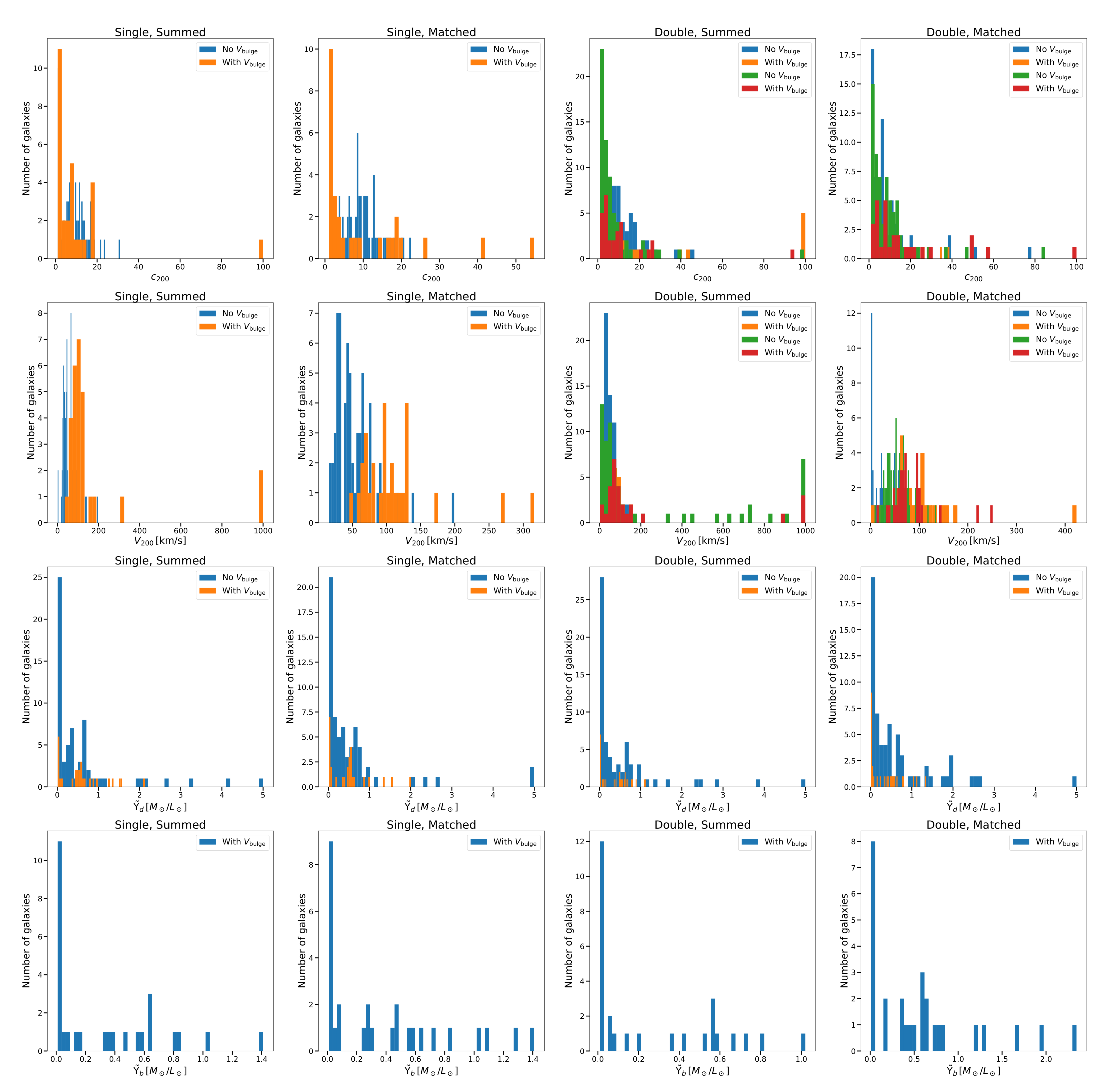}
}
\caption{\textbf{Particle masses fixed ($m_1 = 10^{1.5}\,m_{22}$, $m_2 = 10^{1.8}\, m_{22}$): }Distributions for $c_{200}$ (top row), $V_{200}$ (2nd from top row), $\tilde{\Upsilon}_d$ (third from top row), and $\tilde{\Upsilon}_b$ (bottom row) for the single, summed (left column), single, matched (2nd column from the left), double, summed (3rd column from the left), and double, matched (right column) models.}
\label{fig:psi_mfix_params_dist_ex}
\end{figure}

\begin{figure}
\centering
\makebox[0pt]{
\includegraphics[width=0.7\paperwidth]{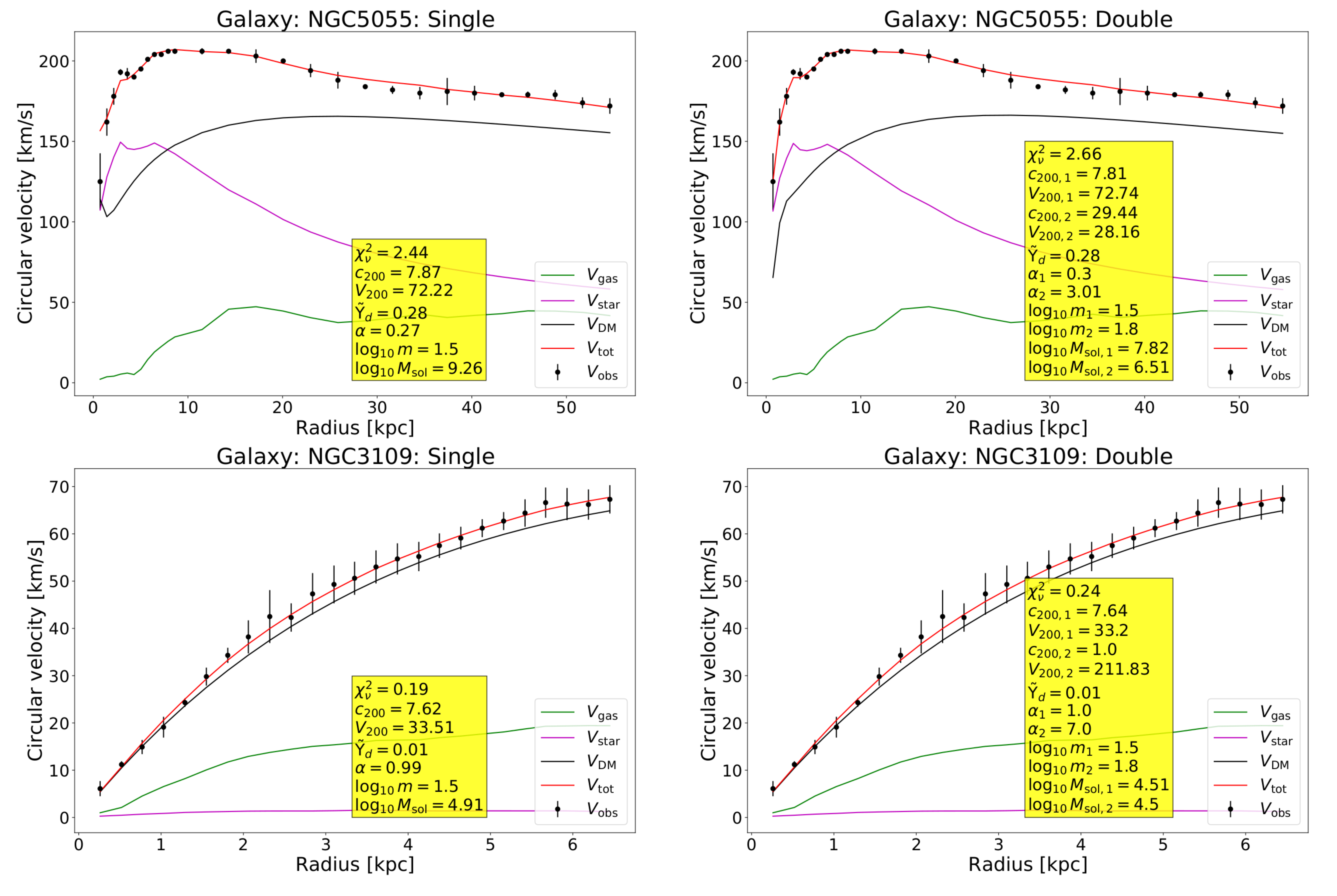}
}
\caption{\textbf{Particle masses fixed ($m_1 = 10^{1.5}\,m_{22}$, $m_2 = 10^{1.8}\, m_{22}$): }Rotation curves for galaxies NGC5055 (top row) and NGC3109 (bottom row) for the assumed models: single, summed (left); double, summed (right).}
\label{fig:psi_Summed_mfix_rotcurves_ex}
\end{figure}

\begin{figure}
\centering
\makebox[0pt]{
\includegraphics[width=0.7\paperwidth]{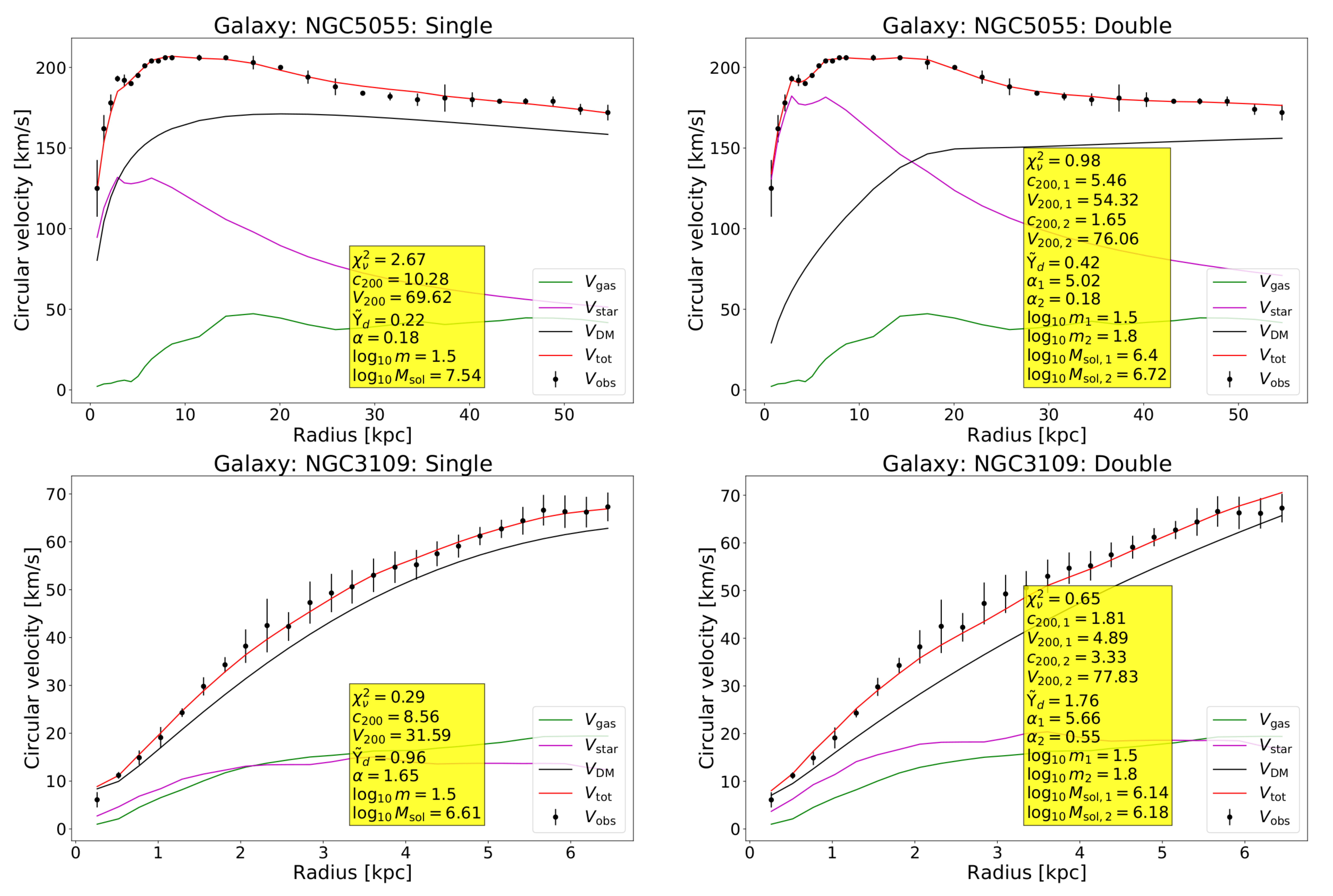}
}
\caption{\textbf{Particle masses fixed ($m_1 = 10^{1.5}\,m_{22}$, $m_2 = 10^{1.8}\, m_{22}$): }Rotation curves for galaxies NGC5055 (top row) and NGC3109 (bottom row)  for the assumed models: single, matched (left); double, matched (right).}
\label{fig:psi_Matched_mfix_rotcurves_ex}
\end{figure}

\begin{figure}
\centering
\makebox[0pt]{
\includegraphics[width=0.85\paperwidth]{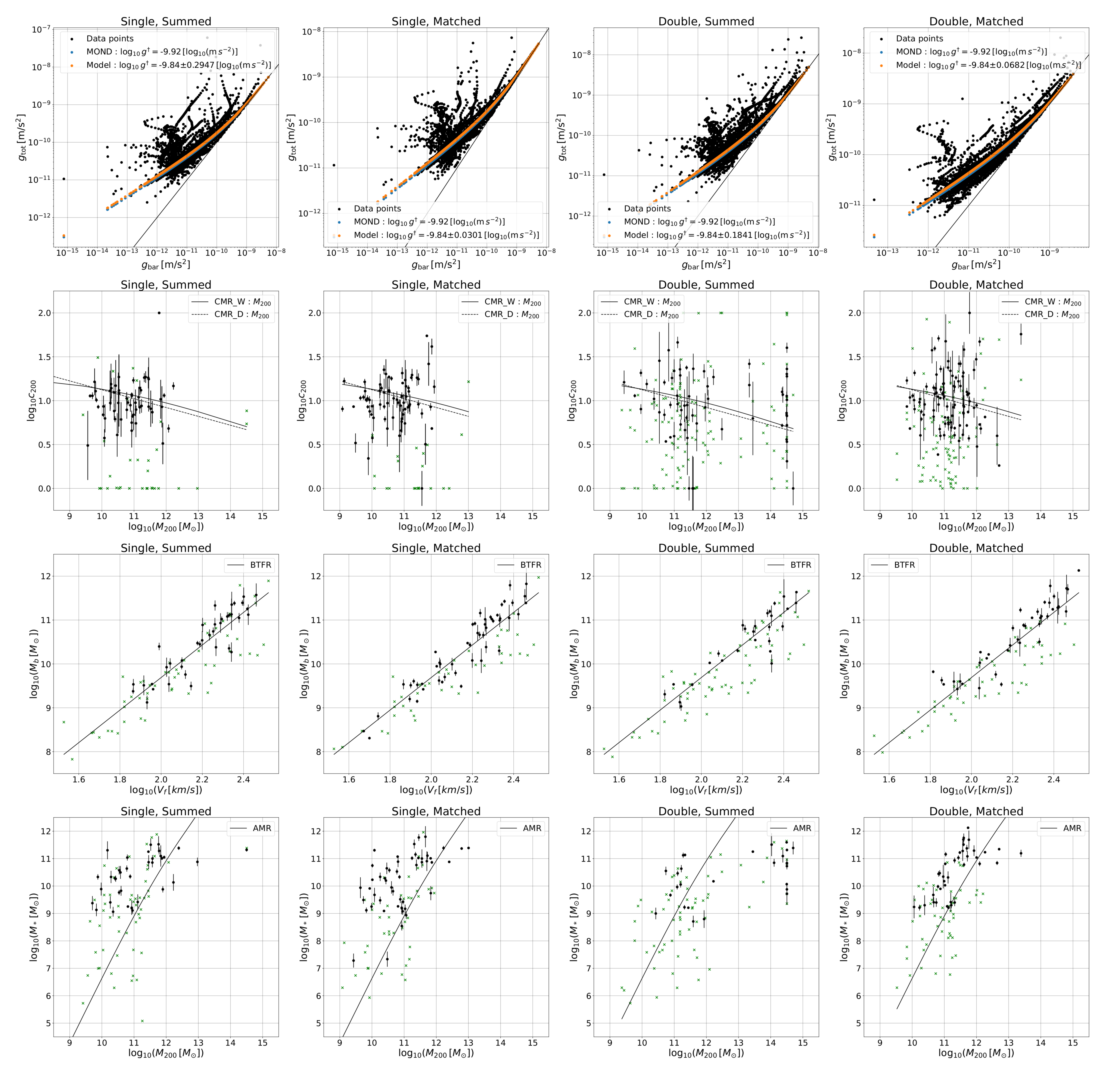}
}
\caption{\textbf{Particle masses fixed ($m_1 = 10^{1.5}\,m_{22}$, $m_2 = 10^{1.8}\, m_{22}$): }Empirical relations for the single, summed (left column), single, matched (2nd column from the left), double, summed (3rd column from the left), and double, matched (right column) models.  \textbf{Top row: }Gravitational RARs.  Blue points correspond to Eq. (\ref{eq:grar}) for the MOND value $g_\dagger=1.2\times10^{-10}\,\text{m}/\text{s}^2$, and orange points to Eq. (\ref{eq:grar}) with the best fit $g_\dagger$.  \textbf{2nd row from the top: }$\log_{10}c_{200}$ vs. $\log_{10}M_{200}$.  Black solid lines correspond to values calculated from Eq. (\ref{eq:conc_mass_relation_Wa}) and black dashed lines to values calculated from Eq. (\ref{eq:conc_mass_relation_Du}).  \textbf{3rd row from the top: }$\log_{10}M_\text{baryons}$ (i.e. $M_* + M_\text{gas}$) vs. $\log_{10}V_f$.  Black solid lines correspond to values calculated from Eq. (\ref{eq:BTFR}).  \textbf{Bottom row: }$\log_{10}M_*$ vs. $\log_{10}M_{200}$.  Black solid lines correspond to values calculated from Eq. (\ref{eq:AMR}).  For the 2nd from the top, 3rd from the top and bottom rows, the points are marked the same as in Fig. \ref{fig:psi_mfree_MsolvsMhalo}.}
\label{fig:psi_mfix_relations}
\end{figure}

We now show more results for which the particle masses are fixed to $m_1 = 10^{1.5} \, m_{22}$ and $m_2 = 10^{1.8} \, m_{22}$, which gave the best fitting results for the galaxies analyzed.  For the main results see Sec. \ref{sec:results} (Figs. \ref{fig:psi_mfix_BICvsBIC_ex}, \ref{fig:psi_mfix_chivschi_ex}, and  \ref{fig:psi_mfix_MsolvsMhalo_ex}):
\begin{itemize}
\item Fig \ref{fig:psi_mfix_dist_ex} shows the reduced chi-square values;
\item Fig \ref{fig:psi_mfix_params_dist_ex} shows some of the parameter distributions. The 2nd from the bottom and bottom rows show the distributions of $\tilde{\Upsilon}_d$ and $\tilde{\Upsilon}_b$, respectively.  Both distributions tend to peak near the lower boundary of 0.01;
\item Fig. \ref{fig:psi_Summed_mfix_rotcurves_ex} shows rotation curves for the galaxies NGC5055 and NGC3109 for the summed models;
\item Fig. \ref{fig:psi_Matched_mfix_rotcurves_ex} shows rotation curves for the galaxies NGC5055 and NGC3109 for the matched models;
\item Fig. \ref{fig:psi_mfix_relations} shows the empirical relations analyzed (the gravitational RAR of \cite{2016}, the CMR \cite{Dutton:2014xda,Wang:2019ftp}, the BTFR \cite{2015}, and the AMR \cite{2013,2012}).  For the gravitation RAR, all ULDM models give a value of $g^\dagger$ that is close to the MOND value.  All models tend to give significant scatter around both CMR relations, around the BTFR relation, and around the AMR relation, with less scatter around the BTFR relation.
\end{itemize}

\begin{figure}
\centering
\makebox[0pt]{
\includegraphics[width=0.35\paperwidth]{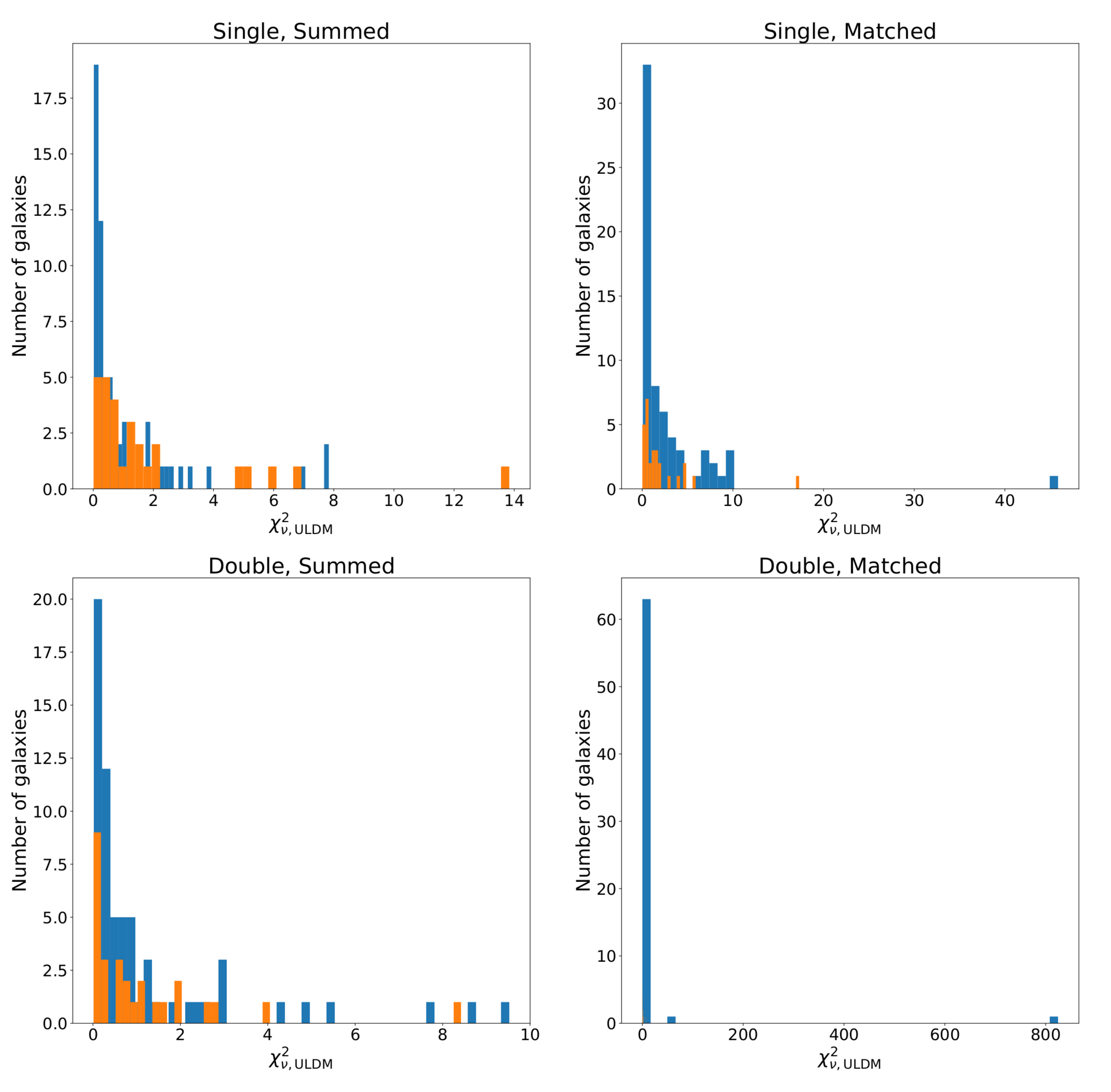}
}
\caption{\textbf{Particle mass free: }Reduced chi-square $\chi^2/\nu$ for the assumed models: single, summed (top, left); single, matched (top, right); double, summed (bottom, left); double, matched (bottom, right).}
\label{fig:psi_mfree_dist}
\end{figure}

\begin{figure}
\centering
\makebox[0pt]{
\includegraphics[width=0.55\paperwidth]{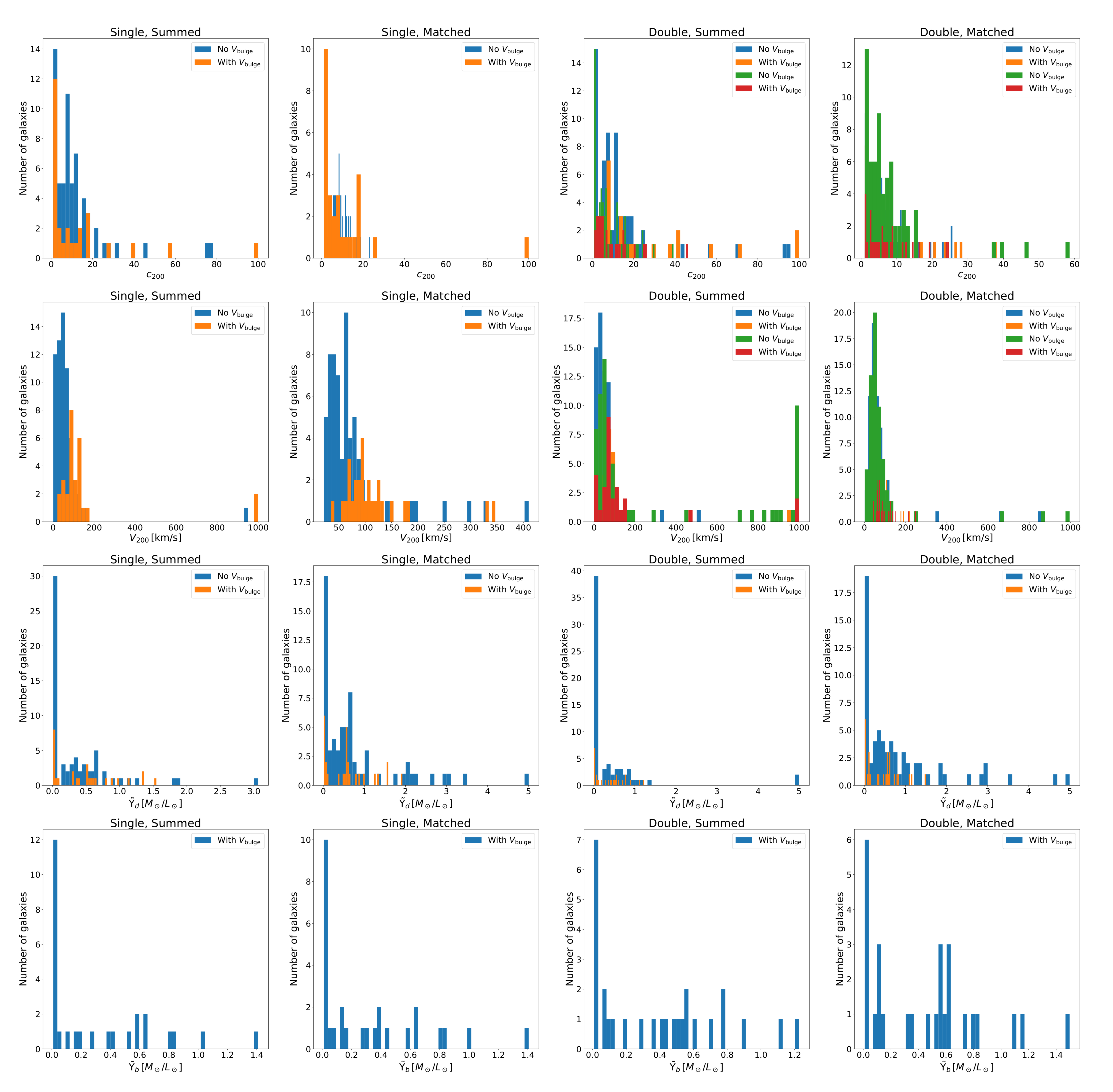}
}
\caption{\textbf{Particle mass free: }Distributions for $c_{200}$ (top row), $V_{200}$ (2nd from top row), $\tilde{\Upsilon}_d$ (third from top row), and $\tilde{\Upsilon}_b$ (bottom row) for the single flavor summed (left column), single flavor matched (2nd column from the left), double flavor summed (3rd column from the left), and double flavor matched (right column) models.}
\label{fig:psi_mfree_params_dist}
\end{figure}

\begin{figure}
\centering
\makebox[0pt]{
\includegraphics[width=0.7\paperwidth]{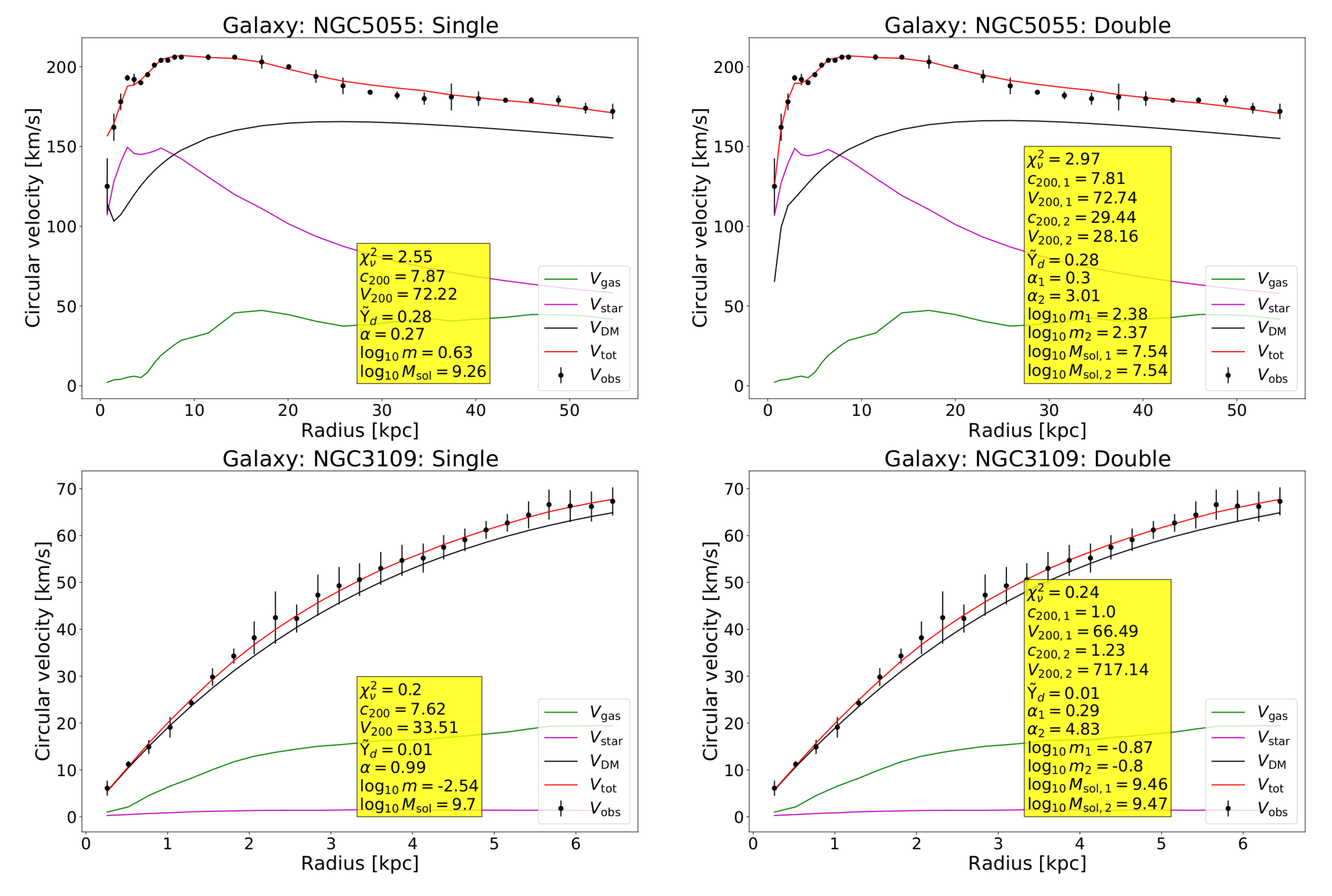}
}
\caption{\textbf{Particle mass free: }Rotation curves for galaxies NGC5055 (top row) and NGC3109 (bottom row) for the assumed models: single, summed (left); double, summed (right).}
\label{fig:psi_Summed_mfree_rotcurves}
\end{figure}

\begin{figure}
\centering
\makebox[0pt]{
\includegraphics[width=0.7\paperwidth]{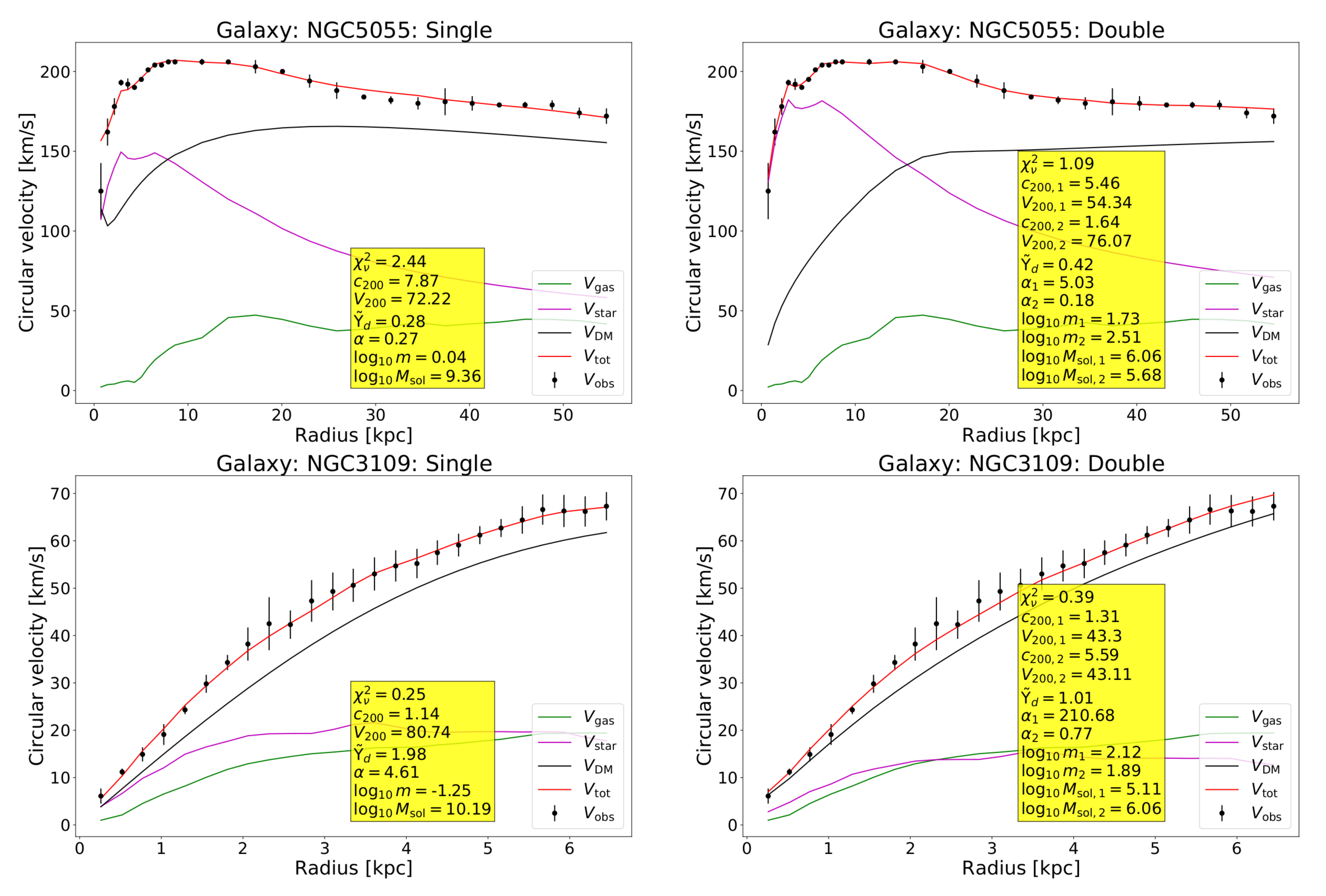}
}
\caption{\textbf{Particle mass free: }Rotation curves for galaxies NGC5055 (top row) and NGC3109 (bottom row)  for the assumed models: single, matched (left); double, matched (right).}
\label{fig:psi_Matched_mfree_rotcurves}
\end{figure}

\begin{figure}
\centering
\makebox[0pt]{
\includegraphics[width=0.85\paperwidth]{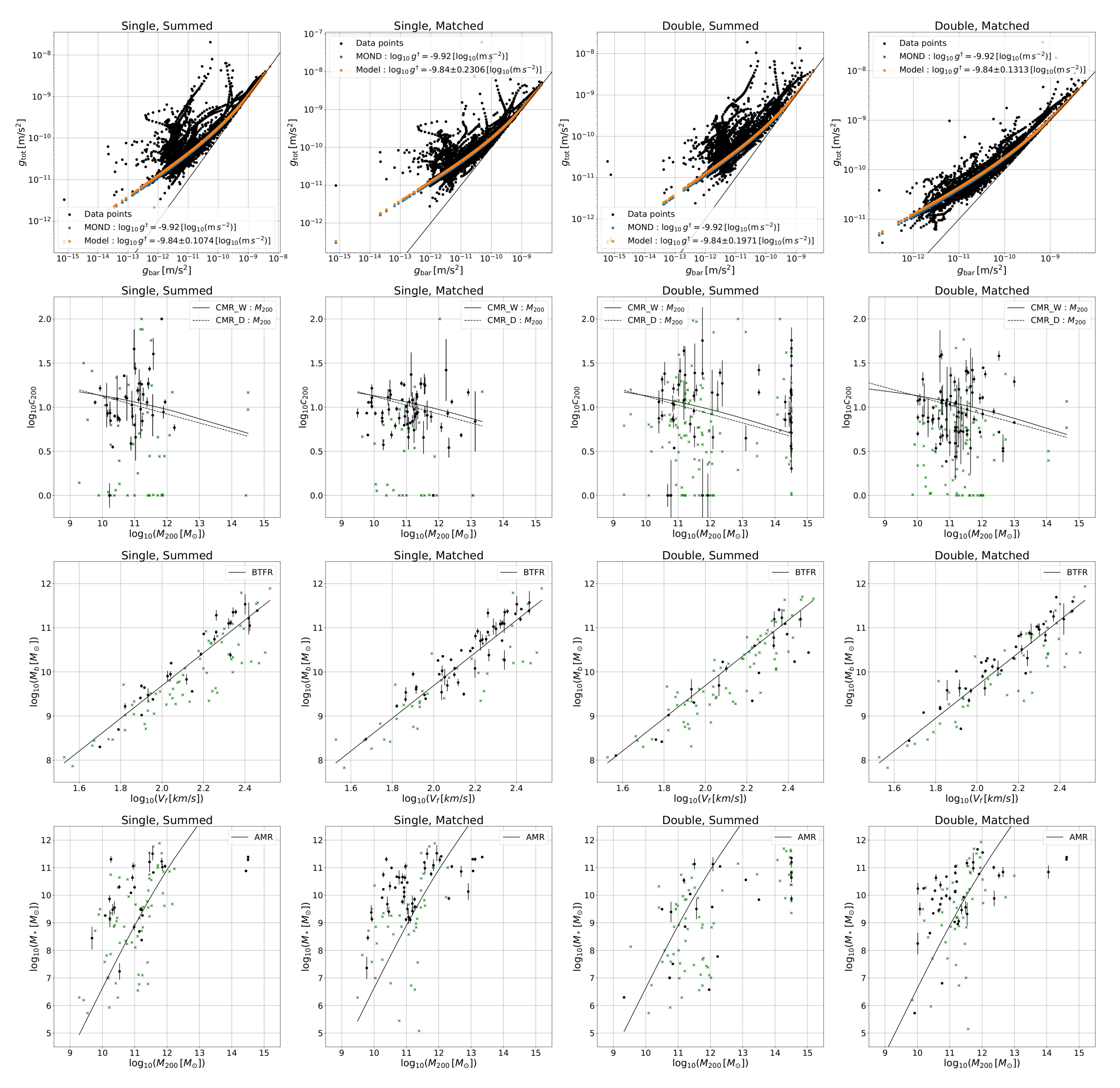}
}
\caption{\textbf{Particle mass free: }Empirical relations for the single, summed (left column), single, matched (2nd column from the left), double, summed (3rd column from the left), and double, matched (right column) models.  \textbf{Top row: }Gravitational RARs.  Blue points correspond to Eq. (\ref{eq:grar}) for the MOND value $g_\dagger=1.2\times10^{-10}\,\text{m}/\text{s}^2$, and orange points to Eq. (\ref{eq:grar}) with the best fit $g_\dagger$.  \textbf{2nd row from the top: }$\log_{10}c_{200}$ vs. $\log_{10}M_{200}$.  Black solid lines correspond to values calculated from Eq. (\ref{eq:conc_mass_relation_Wa}) and black dashed lines to values calculated from Eq. (\ref{eq:conc_mass_relation_Du}).  \textbf{3rd row from the top: }$\log_{10}M_\text{baryons}$ (i.e. $M_* + M_\text{gas}$) vs. $\log_{10}V_f$.  Black solid lines correspond to values calculated from Eq. (\ref{eq:BTFR}).  \textbf{Bottom row: }$\log_{10}M_*$ vs. $\log_{10}M_{200}$.  Black solid lines correspond to values calculated from Eq. (\ref{eq:AMR}).  For the 2nd from the top, 3rd from the top and bottom rows, the points are marked the same as in Fig. \ref{fig:psi_mfree_MsolvsMhalo}.}
\label{fig:psi_mfree_relations}
\end{figure}

We also show more results for which the particle mass is free to vary in the fitting procedure.  For the main results see Sec. \ref{sec:results} (Figs. \ref{fig:psi_mfree_BICvsm}, \ref{fig:psi_mfree_chivschi}, and \ref{fig:psi_mfree_MsolvsMhalo}):
\begin{itemize}  
\item Fig \ref{fig:psi_mfree_dist} shows the reduced chi-square values;
\item Fig \ref{fig:psi_mfree_params_dist} shows some of the parameter distributions.  The 2nd from the bottom and bottom rows show the distributions of $\tilde{\Upsilon}_d$ and $\tilde{\Upsilon}_b$, respectively.  As in Fig. \ref{fig:psi_mfix_params_dist_ex} both distributions tend to peak near the lower boundary of 0.01. 
\item Fig. \ref{fig:psi_Summed_mfree_rotcurves} shows rotation curves for the galaxies NGC5055 and NGC3109 for the summed models;
\item Fig. \ref{fig:psi_Matched_mfree_rotcurves} shows rotation curves for the galaxies NGC5055 and NGC3109 for the matched models;
\item Fig. \ref{fig:psi_mfree_relations} shows the empirical relations analyzed (the gravitational RAR of \cite{2016}, the CMR \cite{Dutton:2014xda,Wang:2019ftp}, the BTFR \cite{2015}, and the AMR \cite{2013,2012}).  As in Fig. \ref{fig:psi_mfix_relations}, all ULDM models give a value of $g^\dagger$ that is close to the MOND value.  All models tend to give significant scatter around both CMR relations, around the BTFR relation, and around the AMR relation, with less scatter around the BTFR relation.
\end{itemize}
\clearpage

\clearpage

\subsection{CDM}\label{sec:app_CDM}
\subsubsection{Comparison with previous CDM studies}\label{sec:app_comp_study}
First, we discuss the reproduction of results from previous studies.  Similarly to \cite{10.1093/mnras/stw3101}, we compare the NFW and DC14 profile fits for the SPARC galaxies.  Here, we take only galaxies with inclinations greater than $30^o$, and we omit galaxies with quality flags equal to three.   This leaves us with 149 galaxies analyzed, rather than the 147 galaxies analyzed  in \cite{10.1093/mnras/stw3101}.  We compare our results to the flat prior analysis of \cite{10.1093/mnras/stw3101}, where the free parameters are constrained to the ranges $1 \leq c_{200} \leq 100$, $10 \leq V_{200} \leq 500$, $0.3 \leq \tilde{\Upsilon}_d \leq 0.8$, and $0.3 \leq \tilde{\Upsilon}_b \leq 0.8$.  Finally, we also take the constraint $(M_* + M_\text{gas})/M_\text{DM} < 0.2$ as in \cite{10.1093/mnras/stw3101}.  With this analysis set up, we confirm that the DC14 profile better fits more galaxies analyzed.  We are able to approximately reproduce Fig. 1 of \cite{10.1093/mnras/stw3101} as well as the rotation curves of Figs. A1-A7 for the flat prior case.  We obtain the median reduced chi-squared (Eq. \ref{eq:chi_square}) for all galaxies to be $\chi^2_{\nu,\text{NFW}} \approx 1.55$ and $\chi^2_{\nu,\text{DC}14} \approx 0.85$.  We also obtain the following fraction of galaxies, $f$, with $\Delta \text{BIC}= \text{BIC}_\text{NFW} - \text{BIC}_\text{DC14}$:  $f = 0.36$ for $\Delta \text{BIC} > 6$; $f = 0.13$ for $6 \geq \Delta \text{BIC} > 2$; $f = 0.28$ for $2 \geq \Delta \text{BIC} > -2$; $f = 0.05$ for $-2 \geq \Delta \text{BIC} > -6$; and $f = 0.17$ for $-6 \geq \Delta \text{BIC}$.  

Next, we discuss the reproduction of results from \cite{Li:2020iib}.  Here, we analyze all 175 galaxies, take uniform priors, and constrain the ranges of parameters to be $1\leq c_{200}\leq 1000$, $10 \leq V_{200} \leq 500$, $0.01\leq \tilde{\Upsilon}_d \leq 5$, $0.01 \leq \tilde{\Upsilon}_b \leq 5$, and $5 \times 10^{-3} \leq \alpha \leq 5$.  We are able to approximately reproduce Fig. 1 and the rotation curve plots for a handful of galaxies (for the Burkert, DC14, Einasto, and NFW profiles).  However, we obtain differing results for the free parameters, especially the mass-to-light ratios (which we take to have uniform rather than log-normal priors).  We also find that many galaxies have a strong degeneracy in the best fit the mass-to-light ratio, and that many values of the mass-to-light ratio can result in similar maximum likelihoods.  This could also be a contributing factor to the differences in the best fit mass-to-light ratios obtained.  Most importantly, we find that cored profiles (Burkert, DC14, and Einasto) better fit, in general, the SPARC catalog galaxies, while the Einasto profile tends to have the best fit values for the reduced chi-square.

Finally, we discuss the reproduction of the results from \cite{2021}, in which, among others, the NFW and Einasto profile fits for the SPARC galaxies are compared.  Here, we take only galaxies with a total number of measured circular velocites $N \geq 10$ and a quality flag less than three, leaving a total of 121 galaxies analyzed as in \cite{2021}.  We take uniform priors and constrain the ranges of parameters to be $1 \leq c_{200} \leq 100$, $1 \leq V_{200} \leq 500$, $0.01 \leq \tilde{\Upsilon}_d \leq 5$, and $10^{-3} \leq \alpha \leq 10$.  We also take $\tilde{\Upsilon}_b = 1.4 \tilde{\Upsilon}_d$ and minimize the chi-squared given by,
\begin{align}
\chi^2_\Upsilon = \left(\frac{\tilde{\Upsilon}_d - \overline{\Upsilon}_d}{\sigma_{\Upsilon_d}}\right)^2 + \chi^2,
\end{align}
where $\chi^2$ is given by Eq. (\ref{eq:chi_square}).  With this analysis, we confirm that the Einasto profile gives a reduced chi-square closer to one than the NFW profile for many of the galaxies analyzed.  We are able to approximately reproduce Figs. 1, 6, and 11 (for the Einasto and NFW profiles) of \cite{2021}.  We find the mean and median reduced chi-squared for all galaxies  analyzed to be $\chi^{2,\text{median}}_{\nu,\text{NFW}} = 1.44$, $\chi^{2,\text{mean}}_{\nu,\text{NFW}} = 3.14$, $\chi^{2,\text{median}}_{\nu,\text{Einasto}} = 0.78$, and $\chi^{2,\text{mean}}_{\nu,\text{Einasto}} = 1.69$.

\subsubsection{Main CDM results} \label{sec:app_CDM_main_results}
We begin with comparing the use of different priors for each of the CDM models.  We perform maximum likelihood estimates for 120 galaxies in the SPARC catalog with inclinations greater than $30^o$, quality flags not equal to three, and nonzero SPARC measurements for the maximum circular velocity $V_f$.  These galaxies also have a total number of circular velocity measurements greater than four (for galaxies without bulge components) or greater than five (for galaxies with bulge components), which is the total number of parameters for the Einasto model.  The fits are performed for five different cases:  (1) uniform priors on all parameters with the possible parameter ranges discussed previously; (2) case (1) with the change $0 < c_{200} < \infty$; (3) case (1) with the change $0 < V_{200} < \infty$; (4) case (1) with the change $0 < \tilde{\Upsilon}_d < \infty$; (5) case (1) with the change $0 < \tilde{\Upsilon}_b < \infty$.

\begin{figure}
\centering
\makebox[0pt]{
\includegraphics[width=0.75\paperwidth]{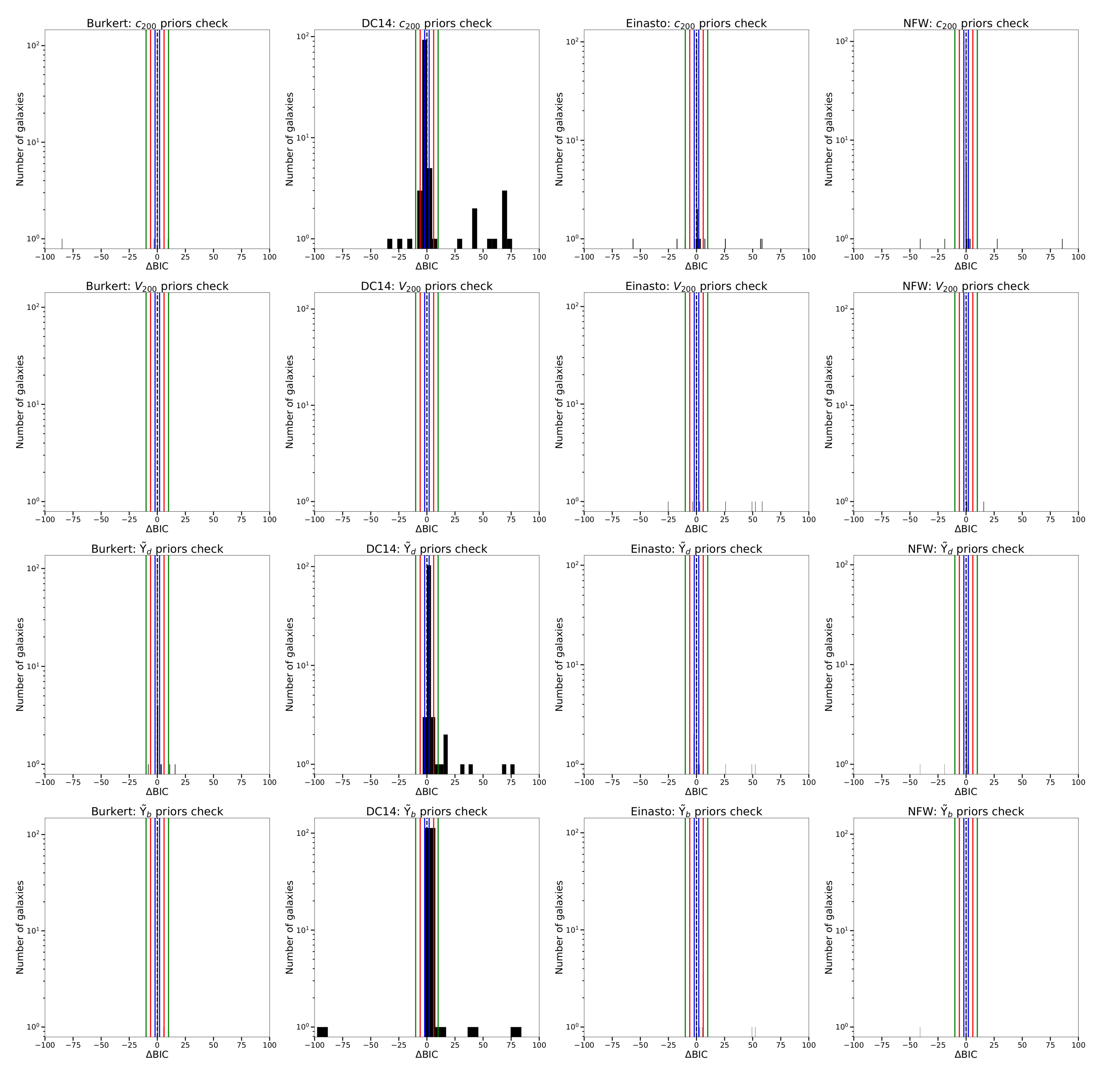}
}
\caption{Distributions of the difference between BIC statistics for different priors cases.  We check five different priors cases:  (1) uniform priors and finite parameter ranges for all parameters; (2) case (1) with the change $0 < c_{200} < \infty$; (3) case (1) with the change $0 < V_{200} < \infty$; (4) case (1) with the change $0 < \tilde{\Upsilon}_d < \infty$; (5) case (1) with the change $0 < \tilde{\Upsilon}_b < \infty$.  The difference between BIC statistics is given by as $\Delta \text{BIC} = \text{BIC}_{(1)} - \text{BIC}_{(i)}$, where $i=2$ corresponds to the top row, $i=3$ to the 2nd row from the top, $i=4$ to the third row from the top, and $i=5$ to the bottom row.  The Burkert model corresponds to the left column, the DC14 model to the 2nd column from the left, the Einasto model to the 3rd column from the left, and the NFW model to the right column.}
\label{fig:CDM_prior_BICdiffs}
\end{figure}

Fig. \ref{fig:CDM_prior_BICdiffs} shows the distributions of the difference in the BIC statistics for each of the SPARC galaxies analyzed.  Each of the prior cases (2)-(5) are compared to prior case (1) with the definition $\Delta \text{BIC} = \text{BIC}_{(1)} - \text{BIC}_{(i)}$, where $i=2$ corresponds to the top row, $i=3$ to the 2nd row from the top, $i=4$ to the third row from the top, and $i=5$ to the bottom row.  The Burkert model corresponds to the left column, the DC14 model to the 2nd column from the left, the Einasto model to the 3rd column from the left, and the NFW model to the right column.  The DC14 model should be viewed differently from the other models for this particular test of different prior cases.  This is due to the fact that one of the parameters in the DC14 model, $V_{200}$, is constrained from $\log_{10}\left(M_*/M_\text{halo}\right) < - 1.3$.  Therefore, for prior case (2), $c_{200}$ is constrained to a finite parameter range due to the constraint on $V_{200}$, and vice versa for prior case (3).  Nonetheless, the DC14 model seems to have the strongest dependence on the parameter ranges chosen for $c_{200}$ and $\tilde{\Upsilon}_d$ compared to the other models, while there are some outliers for the other models.   We show the rotation curves for some example galaxies that give $\Delta \text{BIC} > 6$ for each model in App. \ref{sec:app_results}.

\begin{figure}
\centering
\makebox[0pt]{
\includegraphics[width=0.75\paperwidth]{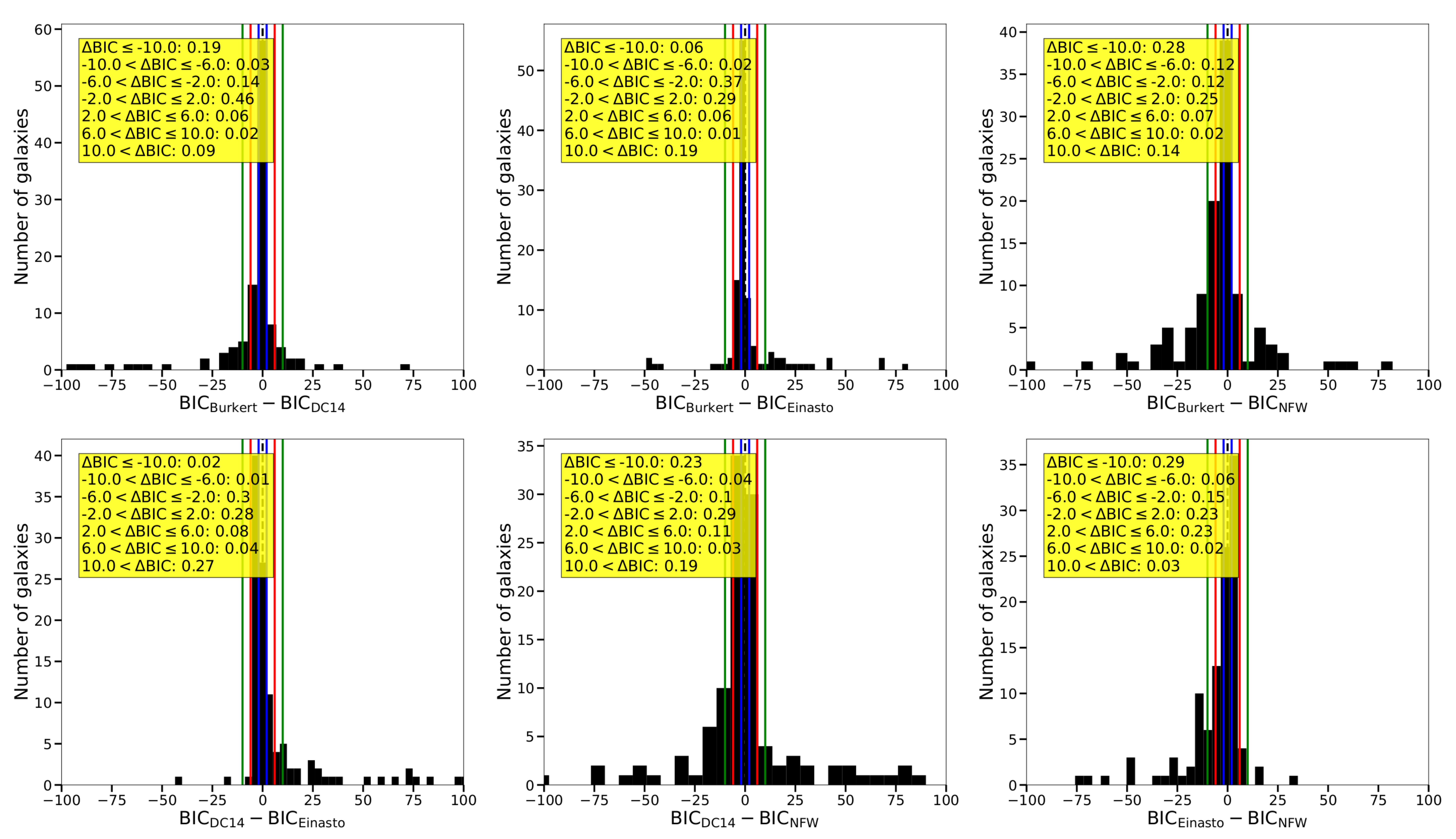}
}
\caption{BIC statistics for:  Burkert vs. DC14 (top left); Burkert vs. Einasto (top middle); Burkert vs. NFW (top right); DC14 vs. Einasto (bottom left); DC14 vs. NFW (bottom middle); and Einasto vs. NFW (bottom right).  Black points correspond to each of the galaxies analyzed, the black dashed line corresponds to $\Delta$BIC$=0$, blue lines correspond to $\left|\Delta\text{BIC}\right|=2$, red lines to $\left|\Delta\text{BIC}\right|=6$, and green lines to $\left|\Delta\text{BIC}\right|=10$. Inset is the fraction of galaxies that fall within a given range for $\Delta \text{BIC}$.}
\label{fig:CDM_BICvsBIC}
\end{figure}

Now, we discuss how the different CDM models compare to each other assuming uniform priors and finite ranges for each free parameter.  Fig. \ref{fig:CDM_BICvsBIC} shows the BIC for each model compared to the BIC for each other model.  The Burkert model is compared to the DC14 model (top left), the Einasto model (top middle), and the NFW model (top right), the DC14 model is compared to the Einasto model (bottom left) and to the NFW model (bottom middle).  Finally, the Einasto model is compared to the NFW model (bottom right).  The lines for $\Delta\text{BIC}=0$, $\left|\Delta\text{BIC}\right|=2$, $\left|\Delta\text{BIC}\right|=6$, and $\left|\Delta\text{BIC}\right|=10$ are displayed as the black dashed, blue, red, and green lines.  The fraction of galaxies that fall within a particular range for $\Delta \text{BIC}$ is shown in the insets.  Both the Burkert and Einasto models tend to perform better than the DC14 and NFW models.  However, almost half of the galaxies analyzed show no preference for the either the Burkert or DC14 models.  A significant portion of galaxies show mild evidence for the Burkert model over the Einasto model, while another significant portion shows decisive evidence for the Einasto model.

\begin{figure}
\centering
\makebox[0pt]{
\includegraphics[width=0.35\paperwidth]{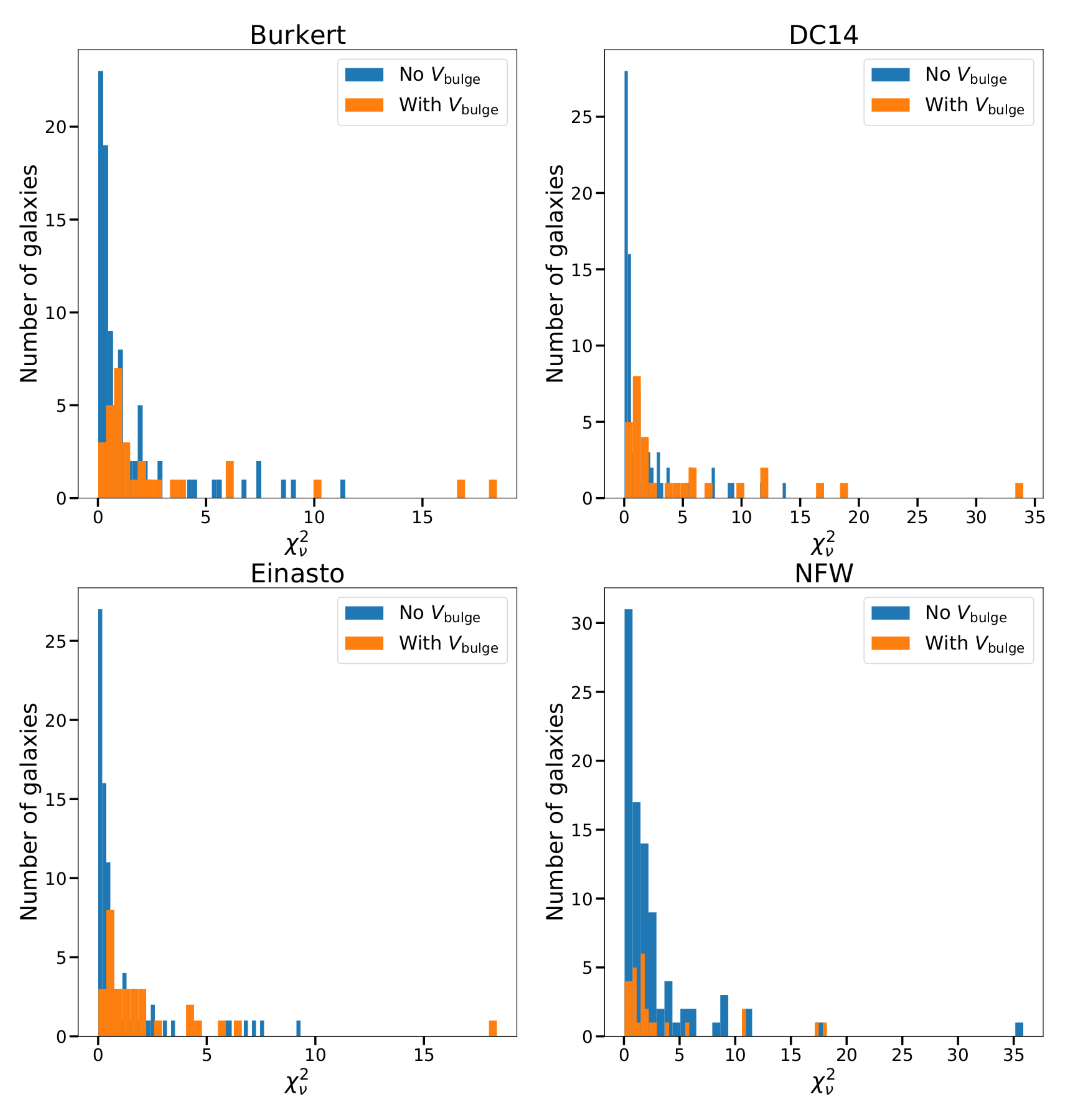}
}
\caption{Reduced chi-square $\chi^2/\nu$ (top row) and BIC (bottom row) for the Burkert (left column), DC14 (2nd from left column), Einasto (third from left column), and the NFW (right column) models.}
\label{fig:CDM_dists}
\end{figure}

\begin{figure}
\centering
\makebox[0pt]{
\includegraphics[width=0.55\paperwidth]{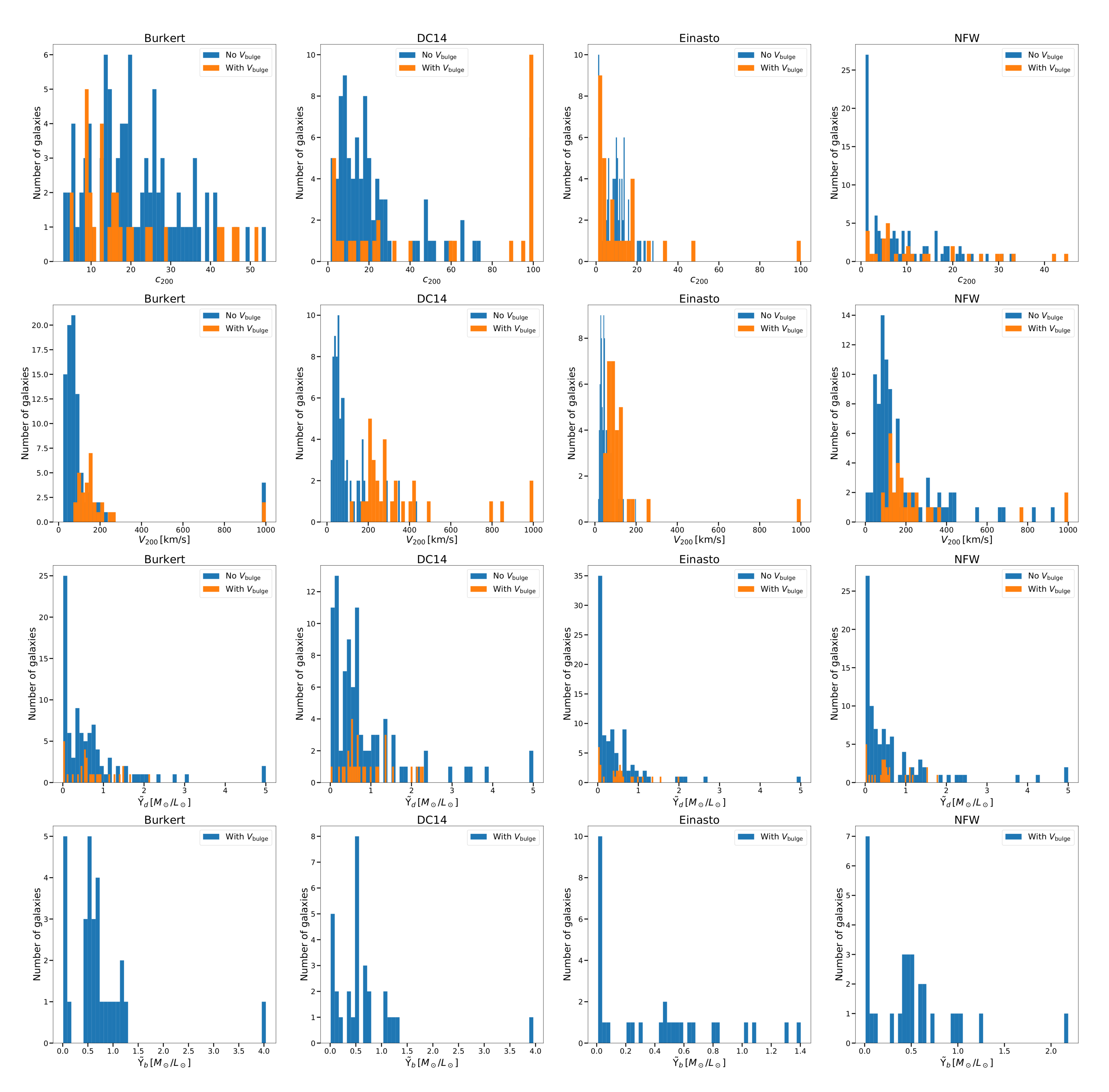}
}
\caption{Distributions for $c_{200}$ (top row), $V_{200}$ (2nd from top row), $\tilde{\Upsilon}_d$ (third from top row), and $\tilde{\Upsilon}_b$ (bottom row) for the Burkert (left column), DC14 (2nd from left column), Einasto (third from left column), and the NFW (right column) models.}
\label{fig:CDM_params_dist}
\end{figure}

\begin{figure}
\centering
\makebox[0pt]{
\includegraphics[width=0.85\paperwidth]{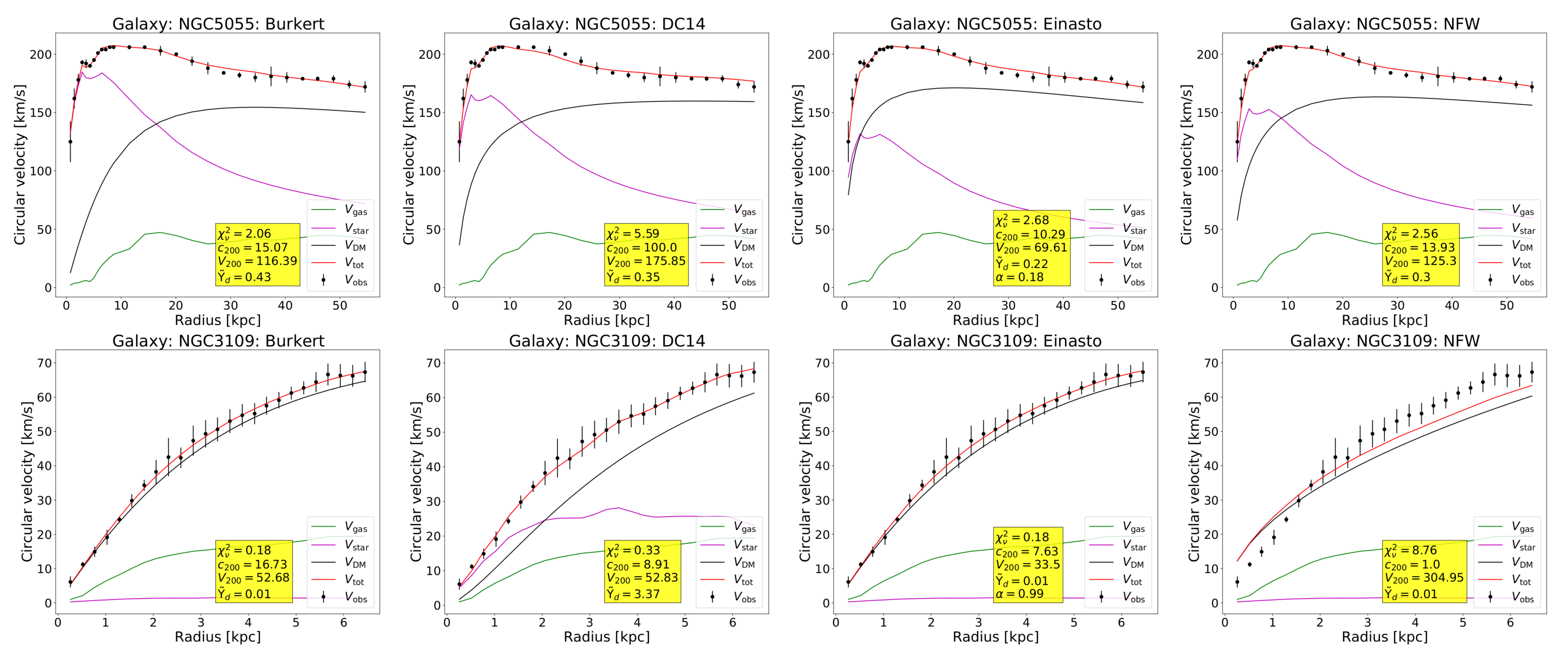}
}
\caption{Rotation curves for galaxies NGC5055 (top row) and NGC3109 (bottom row) for the Burkert (left column), DC14 (2nd from left column), Einasto (third from left column), and the NFW (right column) models.}
\label{fig:CDM_rotcurves}
\end{figure}

\begin{figure}
\centering
\makebox[0pt]{
\includegraphics[width=0.5\paperwidth]{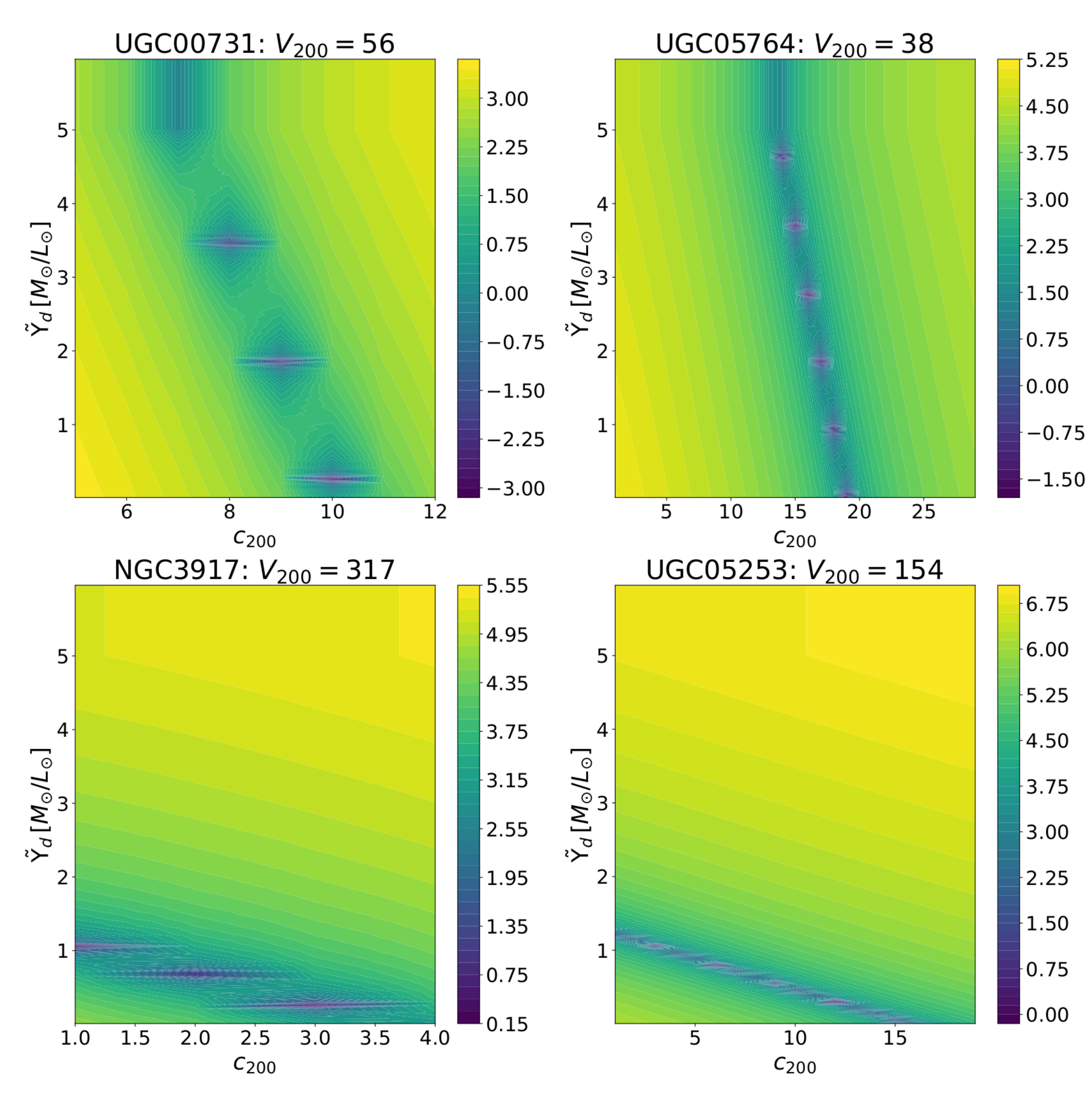}
}
\caption{Galaxies for which the best fit $\tilde{\Upsilon}_d$ is strongly degenerate (top row) and those for which the degeneracy is weaker (bottom row).}
\label{fig:CDM_c200_MLd_chisq_cont}
\end{figure}

\begin{figure}
\centering
\makebox[0pt]{
\includegraphics[width=0.85\paperwidth]{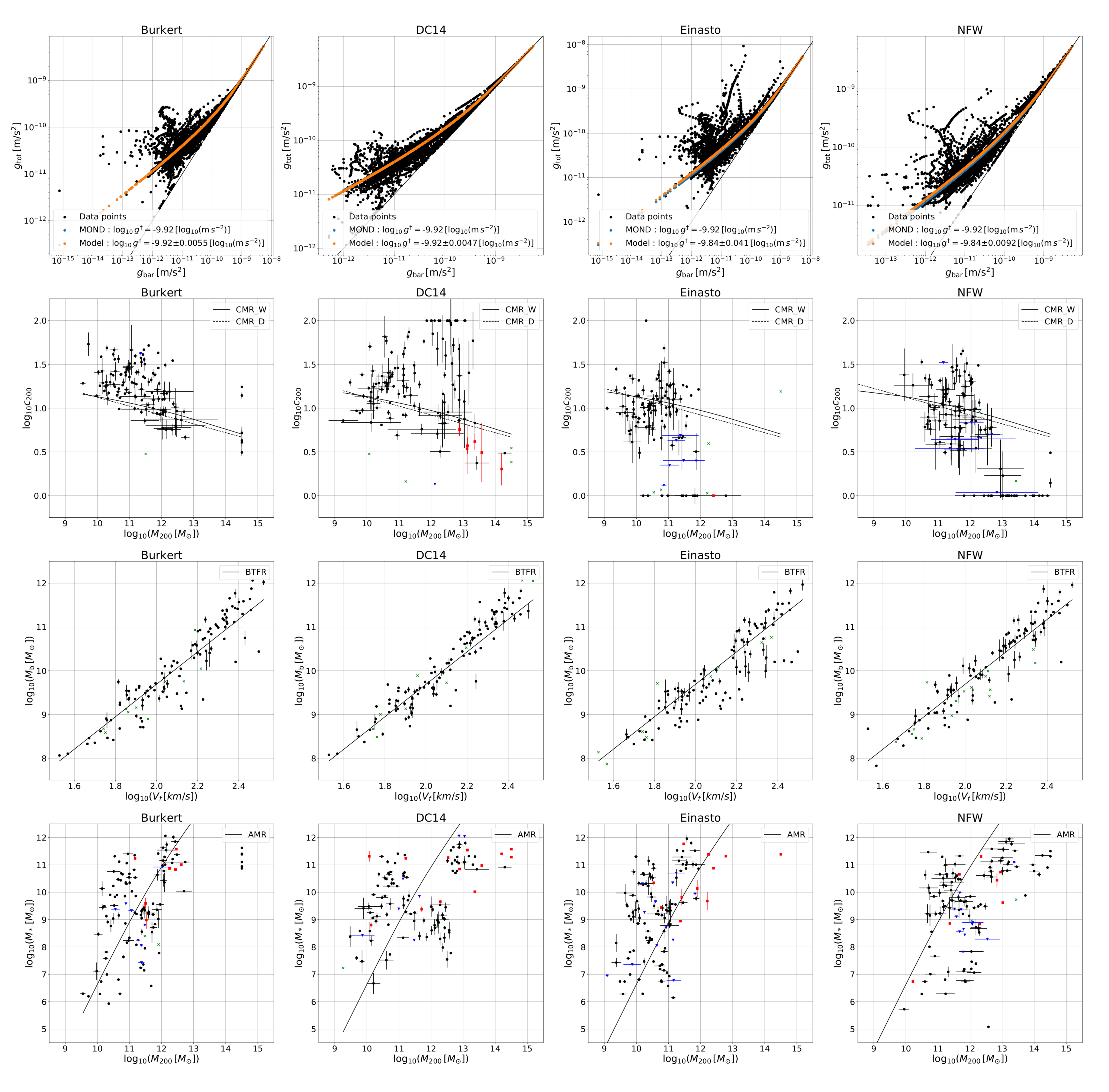}
}
\caption{Empirical relations for the Burkert (left column), DC14 (2nd column from the left), Einasto (3rd column from the left), and NFW (right column) models.  \textbf{Top row: }Gravitational RARs.  Blue points correspond to Eq. (\ref{eq:grar}) for the MOND value $g_\dagger=1.2\times10^{-10}\,\text{m}/\text{s}^2$, and orange points to Eq. (\ref{eq:grar}) with the best fit $g_\dagger$.  \textbf{2nd row from the top: }$\log_{10}c_{200}$ vs. $\log_{10}M_{200}$.  Black solid lines correspond to values calculated from Eq. (\ref{eq:conc_mass_relation_Wa}) and black dashed lines to values calculated from Eq. (\ref{eq:conc_mass_relation_Du}).  \textbf{3rd row from the top: }$\log_{10}M_\text{baryons}$ (i.e. $M_* + M_\text{gas}$) vs. $\log_{10}V_f$.  Black solid lines correspond to values calculated from Eq. (\ref{eq:BTFR}).  \textbf{Bottom row: }$\log_{10}M_*$ vs. $\log_{10}M_{200}$.  Black solid lines correspond to values calculated from Eq. (\ref{eq:AMR}).  For the 2nd from the top, 3rd from the top and bottom rows, the points are marked the same as in Fig. \ref{fig:psi_mfree_MsolvsMhalo}.}
\label{fig:CDM_relations}
\end{figure}

Finally, we show the resulting statistical and parameter distributions, example rotation curves, and empirical relations for the CDM models:
\begin{itemize}  
\item Fig. \ref{fig:CDM_dists} shows the reduced chi-square values;
\item Fig. \ref{fig:CDM_params_dist} shows the parameter distributions for each of the CDM models.  The 2nd from the bottom and bottom rows show the distributions of $\tilde{\Upsilon}_d$ and $\tilde{\Upsilon}_b$, respectively.  For each model, the $\tilde{\Upsilon}_d$ distribution tends to peak near the lower boundary of 0.01, while for the Einasto and NFW models, the $\tilde{\Upsilon}_b$ distribution also tends to peak near the lower boundary of 0.01.  For the Burkert, the distribution of $\tilde{\Upsilon}_b$ has equal numbers of galaxies showing preference for the lower boundary of 0.01 and $\approx 0.6$.  For the DC14, the distribution of $\tilde{\Upsilon}_b$ tends to peak around $\approx 0.6$;
\item Fig. \ref{fig:CDM_rotcurves} shows rotation curves for the galaxies NGC5055 and NGC3109 for each of the CDM models;
\item Fig. \ref{fig:CDM_c200_MLd_chisq_cont} shows example galaxies for which the degeneracy in the best fit $\tilde{\Upsilon}_d$ is strong or weak.
\item Fig. \ref{fig:CDM_relations} shows the empirical relations analyzed (the gravitational RAR of \cite{2016}, the CMR \cite{Dutton:2014xda,Wang:2019ftp}, the BTFR \cite{2015}, and the AMR \cite{2013,2012}).  As in Figs. \ref{fig:psi_mfix_relations} and \ref{fig:psi_mfree_relations}, all models give a value of $g^\dagger$ that is close to the MOND value.  All models tend to give significant scatter around both CMR relations, around the BTFR relation, and around the AMR relation, with less scatter around the BTFR relation.
\end{itemize}
\clearpage

\end{appendix}

\end{document}